\documentclass[%
 reprint,
 amsmath,amssymb,
 aps,
prb,
]{revtex4-2}

\usepackage{graphicx}% Include figure files
\usepackage{dcolumn}% Align table columns on decimal point
\usepackage{bm}% bold math
\usepackage{siunitx}
\usepackage{float}
\begin{document}

\preprint{APS/123-QED}

\title{Electronic and structural properties of atomically thin metallenes}

\author{Kameyab Raza Abidi}
\email{kameyab.r.abidi@jyu.fi}
 \author{Pekka Koskinen}%
 \email{Corresponding author, pekka.j.koskinen@jyu.fi}
\affiliation{%
 NanoScience Center, Department of Physics, University of Jyväskylä, 40014 Jyväskylä, Finland
}%

\date{\today}% It is always \today, today,
             %  but any date may be explicitly specified

\begin{abstract}
Although metallic elements favor three-dimensional (3D) geometries due to their isotropic, metallic bonding, experiments have reported metals also with two-dimensional (2D) allotropes, the so-called metallenes. And while bulk metals' electronic and structural properties are well known, the corresponding knowledge for atomically thin metallenes remains scattered. Therefore, in this work, we use density-functional theory to investigate the electronic and structural properties of 45 elemental metals with honeycomb, square, and hexagonal lattices, along with their buckled counterparts, resulting in a comprehensive catalog of 270 metallenes with their properties. We systematically present their structural, energetic, and electronic structure properties and discuss similarities and differences compared to their 3D counterparts. As a result, simple and noble metals exhibit similar characteristics and lack buckled hexagonal lattice. Apart from scattered exceptions, the trends in several properties, such as bond lengths, cohesion energies, and projected densities of states, are governed by coordination numbers and exhibit systematic patterns. This systematic reporting provides a necessary reference for the selection and categorization of metallenes for further experimental efforts to develop them for catalytic, sensing, plasmonic, and nanoelectronics applications.

\end{abstract}

%To streamline the synthesis strategies of metallenes, the knowledge of their electronic structure can help

%\keywords{Suggested keywords}%Use showkeys class option if keyword
                              %display desired
\maketitle

%\tableofcontents

\section{\label{sec:intro}Introduction}
The current understanding of metals' electronic structure mainly relies on the seminal work of Slater \cite{slaterElectronicStructureMetals1934} and Wigner \cite{wignerInteractionElectronsMetals1934}, which has facilitated the investigation of metallic bonding and related phenomena  \cite{ hallElectronicStructureBodyCentred1953, pippardExperimentalAnalysisElectronic1960, lomerElectronicStructureChromium1962, harrisonElectronicStructureSeries1963,lomerElectronicStructurePure1969a, lomerElectronicStructurePure1969, hyghElectronicStructureTitanium1970, andersenElectronicStructureFcc1970, louieElectronicStructureCesium1974,
wooElectronicStructureMetals1975,matsuuraElectronicStructureMetals1975,jepsenElectronicStructureHcp1975, soElectronicStructureMetals1977, soElectronicStructureMetals1977a, laiElectronicStructureMetals1978, wangElectronicStructureMetals1979, skriverCohesionElectronicStructure1982, blahaElectronicStructureFermi1985, tripathiElectronicStructureRhodium1988, blahaElectronicStructureHcp1988, harrisonElectronicStructureProperties1989, fusterElectronicStructureRelated1990, leeElectronicStructureMagnetism1993, liuElectronicStructureSemimetals1995, xieElectronicStructuresProperties2001, schillerElectronicStructureMg2004, iotaElectronicStructureMagnetism2007, brenerElectronicStructureSmall2009, pandaElectronicStructureEquilibrium2012,   konigElectronicPropertiesBismuth2021}. Studies have demonstrated the isotropic nature of metallic bonding, where atoms inherently favor three-dimensional arrangements, as described by the liquid drop or jellium model \cite{perdewLiquiddrop1991, perdewjellium2003}. These models predict the delocalization of electrons, with atoms agglomerating to minimize surface energy and maintain electrostatic stability. Despite such tendencies, several research groups have successfully synthesized metals in their two-dimensional (2D) allotropes \cite{wangStabilitySynthesis2D2020b, taSituFabricationFreestanding2021a, yuTwoDimensionalMetalNanostructures2023, dongAirStableLargeArea2D2024}, with the most recently reported synthesis of goldene \cite{kashiwayaSynthesisGoldeneComprising2024a} and atomically thin zirconium in graphene pore \cite{Zirconen_2024}. These metals' 2D allotropes are called \emph{metallenes}.

%However, plenty of 2D materials have been reported after the groundbreaking discovery of graphene \cite{novoselovElectricFieldEffect2004, bhimanapatiRecentAdvancesTwoDimensional2015, choiRecentDevelopmentTwodimensional2017, linRecentAdvances2D2023}.

Currently, the best way to investigate metallenes' structural and electronic structure properties is to adopt a computational approach by density-functional theory (DFT) simulations. DFT has been indispensable in studying the electronic and structural properties of metals \cite{hohenbergInhomogeneousElectronGas1964a,kohnSelfConsistentEquationsIncluding1965a,kratzerBasicsElectronicStructure2019}. It has revealed many key features of metallic bonding, such as ductility and elasticity of face-centred cubic (fcc) metals, and the origin of transition between metallic and insulating states \cite{sunChemicalInteractionsThat2023,mazzoneSystematicInitioStudy2002, aguayoElasticStabilityElectronic2002, kamranInitioExaminationDuctility2009, chellathuraiElectronicStructureTightbinding2020, naumovOriginTransitionsMetallic2015}. Furthermore, DFT has consistently predicted the existence and feasibility of metallenes \cite{nevalaitaAtlasPropertiesElemental2018, nevalaitaIdealTwodimensionalMetals2018, nevalaitaStabilityLimitsElemental2019, nevalaitaFreestanding2DMetals2020, onoDynamicalStabilityTwodimensional2020, onoComprehensiveSearchBuckled2021, renMagnetismElementalTwodimensional2021, sangolkarDensityFunctionalTheory2022, sangolkarProspectusThicknessDependent2022, sangolkarStructureStabilityProperties2022, abidiElectronicStructureElasticity2023, Gentle}.

Although much research has been done, a comprehensive analysis of their structural and electronic properties remains missing. A comprehensive picture of the properties would help develop metallene synthesis, improve stability, and better comprehend and modify chemical driving forces at the atomic level.
Once understood in a detailed fashion, electronic structure could be engineered in applications related to catalysis \cite{greeleyLECTRONICTRUCTUREATALYSIS2002}, energy storage (e.g., Li-ion and Na-ion batteries)  \cite{molendaElectronicStructureEngineering2017}, and electrocatalysis for hydrogen evolution \cite{xiongElectronicStructuralEngineering2022}. The ultimate goal would be a precise control over the metallene properties once synthesized. 

In this work, we use DFT simulations to systematically investigate the electronic and structural properties of atomically thin elemental metallenes. Our study encompasses 45 metals in six 2D lattices: honeycomb ($hc$), square ($sq$), hexagonal ($hex$) and their buckled counterparts ($bhc$, $bsq$, and $bhex$). To elucidate the intrinsic properties of metallenes, our analysis focuses on ideal 2D lattices, deliberately eschewing symmetry-breaking deformations. 
For each metallene, we calculated and analyzed structural and energetic properties, including coordination numbers, bond lengths, atomic densities, and cohesive energies, as well as electronic structure properties, including (projected) densities of states, electron densities, and electron localization functions. The 3D bulk is often used as a reference. We hope our systematic reporting of fundamental properties would provide a solid foundation for further experimental and theoretical studies of 2D metallenes.

%We conducted pair distribution analysis and unveiled the structural features of metallenes, such as the presence of 8 nearest neighbours in buckled square ($bsq$) and 9 in buckled honeycomb ($bhc$) lattices.
%We have identified a systematic trend in the energetics and electronic structural properties in terms of coordination number. For the majority of our analyses, we used 3D bulk as a benchmark.
%We also found that for Group 1, 2, 11, and 12 metals, the 2D $hex$ lattice remains flat without buckling, indicating that these lattices buckle above the convex hull and require compressive strain. In contrast, $hex$ lattices of most transition and post-transition metals buckle at the convex hull, that is, without strain.
%The $bhc$ lattices were found to be equivalent to bilayer $hex$ lattices across all 45 metals. This fundamental insight into the nature of atomically thin metallenes and their comparison with 3D bulk is foundational for future exploration of their potential applications in nanoelectronics, catalysis, and sensing technologies. 

\section{\label{sec:2}Computational Methods and Analysis}

\subsection{Computational methods}
\begin{figure}[t!]
\centering
\includegraphics[width= \columnwidth]{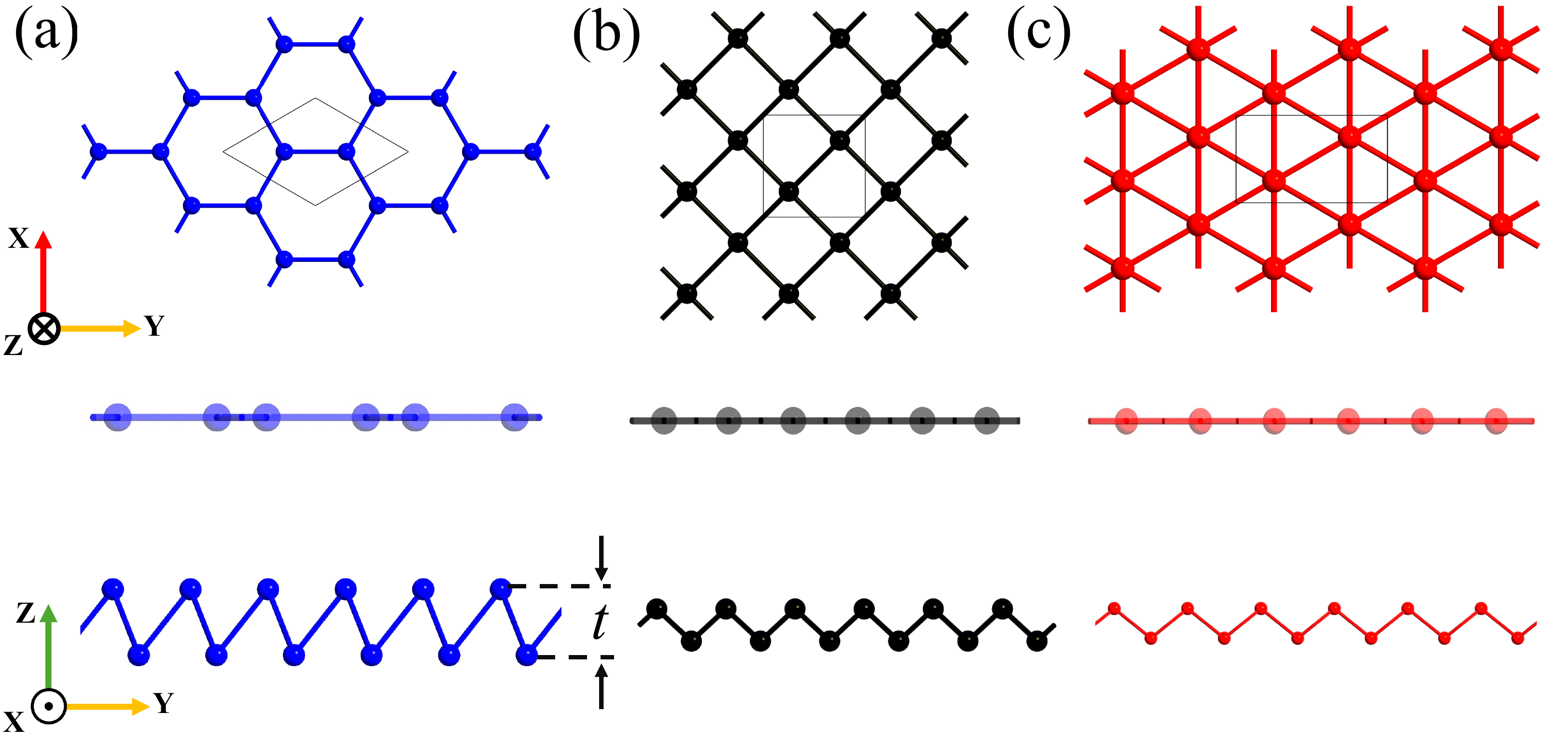}
\caption{Schematic illustrations of (a) honeycomb, (b) square, and (c) hexagonal lattices with top views (upper row) as well as side views of flat lattices (middle row) and buckled lattices (bottom row). The black polygons show the simulation cells. The thickness of the buckled monolayer is \(t\).}
\label{fig:simulation_cell}
\end{figure}

Metallenes were investigated using density-functional theory (DFT) as implemented in QuantumATK (U-2022.12) \cite{smidstrupQuantumATKIntegratedPlatform2020}. The core and valence electrons were modeled using PseudoDojo pseudopotentials \cite{vansettenPseudoDojoTrainingGrading2018}, with valence electrons expanded as a linear combination of atomic orbitals (LCAO) \emph{medium} basis set. Spin-polarized electronic exchange-correlation effects were approximated using the Perdew-Burke-Ernzerhof (PBE) functional \cite{perdewGeneralizedGradientApproximation1996a}. The Brillouin zone was sampled using the Monkhorst-Pack method \cite{monkhorstSpecialPointsBrillouinzone1976a} with $13$ $k$-points in the periodic direction and a vacuum of $20$ $\si{\angstrom}$ along the non-periodic direction. Single-particle states were occupied by the Fermi-Dirac function with $0.05$ $\si{eV}$ energy broadening, and the self-consistency tolerance criterion for the energy was $\leq 10^{-8}$ $\si{eV}$. These parameters are both robust and computationally efficient for modeling metallenes \cite{abidiOptimizingDensityfunctionalSimulations2022a}. All geometries were relaxed using the limited-memory Broyden-Fletcher-Goldfarb-Shanno (LBFGS) algorithm \cite{liuLimitedMemoryBFGS1989} for forces below $10^{-6}$ $\si{eV/\angstrom}$. All computational unit cells had two atoms (Fig.~ \ref{fig:simulation_cell}).

\subsection{Systematic Analysis Tools}
Using the above methods, we conducted electronic and structural analyses as follows:

\paragraph{Energetics and geometrical properties:}
Cohesive energy was calculated as $E_\text{coh}=E_\text{free} - E$, where $E_\text{free}$ is the total energy of a free atom placed in $15$ $\si{\angstrom}$ cube and calculated at $\Gamma$ point, and $E$ is lattice energy per atom. 
To quantify the energy differences between flat and buckled configurations, we introduce a normalized cohesive energy difference, \(\Delta \varepsilon_{coh}^{3D}\), defined as \( (E_{coh}^{buckled} - E_{coh}^{flat}) /{E_{coh}^{3D}}\), where \(E_{coh}^{buckled}\) is cohesive energy of a buckled lattice, \(E_{coh}^{flat}\) is the cohesive energy of the corresponding flat lattice, and \(E_{coh}^{3D}\) is the cohesive energy of 3D bulk. This dimensionless parameter measures the relative stability of buckled structures compared to their flat counterparts. %Normalization by 3D bulk cohesion allows for meaningful comparisons across different elements and lattice geometries, accounting for the inherent differences in cohesive energies among various metallic systems.

All lattice constants and corresponding bond lengths were calculated at relaxed geometries. For buckled lattices, the vertical thickness $t$ equals the separation of atomic planes. Initial guesses $t$ for the buckled geometries were obtained by assuming constant nearest-neighbor distances, implying $t = d\sqrt{|\varepsilon|\,(2-|\varepsilon|)}$, where $\varepsilon$ is the biaxial strain relative to the flat lattice. We also quantified the buckling angle $\alpha$ between the lateral plane and the nearest-neighbor vector. This angle is given by \( \sin\alpha = {t}/{d}\), where \( d \) is the equilibrium bond length. 

\paragraph{Projected density of states (PDOS):}
We calculated the energy averages of the densities of states (DOS) projected for orbital angular momentum quantum number $l$ using the relation
\begin{equation}
 \langle E_{l} \rangle = \frac{\int f(\epsilon) \, \epsilon \, DOS_{l}(\epsilon) \, d\epsilon} {\int f(\epsilon) \, DOS_{l}(\epsilon) \, d\epsilon},
\end{equation}
where $f(\epsilon)$ is the Fermi function, and $DOS_{l}(\epsilon)$ is the projected density of states for the orbital $l$, and $\epsilon$ is the single-particle energy.
The magnitude of $\langle E_{l}\rangle$ reflects the energy level of the orbital relative to the Fermi level, $\epsilon_{f}$, which in our analysis is set to zero. A large negative $\langle E_{l}\rangle$ indicates a deep, more stable orbital, possibly less sensitive to external perturbations. Less negative $\langle E_{l}\rangle$ indicates a shallow state more prone to external perturbations and a greater influence on system properties.
The quantum number $l$ covers orbitals $s$, $p$ ($p_{x}$, $p_{y}$, $p_{z}$), and $d$ ($d_{xy}$, $d_{yz}$, $d_{xz}$, $d_{z^{2}}$, $d_{x^{2}-y^{2}}$), along with their projected combinations such as $xy$ ($p_{x} + p_{y} + d_{xy} + d_{x^{2}-y^{2}}$) and $z$ ($p_{z} + d_{xz} + d_{yz} + d_{z^{2}}$). 
These projections enable a nuanced analysis of the orbital contributions material properties.

\paragraph{Full width at half maximum (FWHM) of electron density:}
The metallene vertical electronic density profiles were analyzed using FWHM metrics. First, we examined the FWHM of the vertical spread of the valence electron density along a vertical line passing through the middle of the interatomic vector \(\vec{R}\), which characterizes the thickness or diffuseness of the electronic cloud. Second, for buckled lattices, the electron density is vertically thicker already due to the buckled geometry. To disentangle the geometric and electronic contributions to the electron density distribution in buckled lattices, we introduced a geometric thickness FWHM$_{gm}$. We calculated this thickness as \(\text{FWHM}_{gm} = {\text{FWHM}_{f}}/{\cos\alpha}\), where $\alpha$ is the buckling angle and \(\text{FWHM}_{f}\) is the FWHM of the corresponding flat lattice at the bond center.
We also computed the onsite thickness \(\text{FWHM}_{onsite}\), which gives the electron cloud thickness at atomic positions.  

These three FWHM analyses provide an adequate scheme to distinguish between structurally dependent and electronically driven differences in the electron density distributions of flat and buckled metallenes. Such analysis is crucial for the rational design and optimization of metallenes for various applications.

%Such analysis is essential for understanding the electronic properties that govern phenomena such as charge transfer, reactivity, and catalytic activity.

\paragraph{Electron localization function (ELF):}
While PDOS offers information about the energy distribution of electronic states and electron density analysis reveals the overall distribution of electronic cloud, the ELF provides complementary spatial information about the degree of electron delocalization, crucial for understanding bonding and metallic character. We calculated the local ELF at bond centers to probe the degree of electron localization or delocalization in these metallic monolayers. The choice of ELF evaluation point is significant as it allows the examination of the electronic structure in the interatomic region and the characterization of the nature of bonding.
%---------------------------- COORDINATION NUMBER --------------------------------------------------

\section{Results and Discussion} 
We will now discuss the structural properties and electronic properties of metallenes as well as their relationships. The aim is to understand structure-property correlations that govern the behavior of these metallenes.

\subsection{Metallene Coordination Numbers}
We begin our analysis by briefly discussing the coordination environment in the six 2D lattices of the 45 elemental metals. While flat 2D lattices exhibit well-defined coordination numbers $CN$ due to their symmetry (3 for $hc$, 4 for $sq$, and 6 for $hex$), buckled structures are more complex. Therefore, the $CN$ analysis helps interpret the subsequent electronic structure results and elucidates trends across the periodic table. 

\begin{figure}[b!]
\centering
\includegraphics[width= \columnwidth]{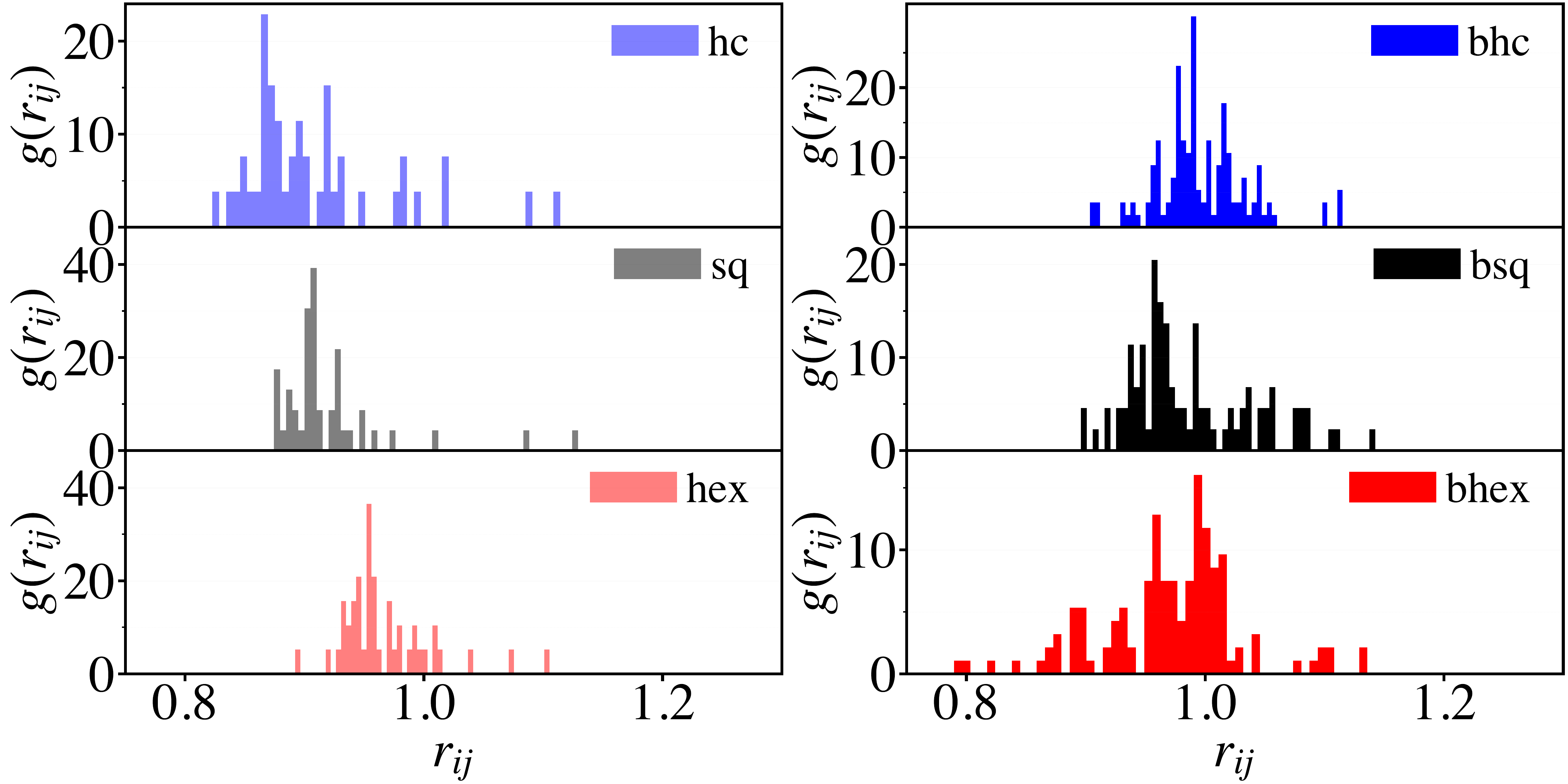}
\caption{The distribution of normalized bond lengths \( r_{ij} \) across the six lattice types of 45 metals. The ordinate shows the number of bond lengths in each bin (bin width = $0.024$).}
\label{fig:pair_dist}
\end{figure}

We employed the pair distribution function (PDF) to unravel the coordination environments of all 2D lattices. It was calculated using the normalized bond length \( r_{ij} \), defined as \( r_{ij} = d_{ij}/d_{nn} \), where \( d_{ij} \) is the distance between atoms $i$ and $j$, and \( d_{nn} \) is the nearest-neighbour distance in 3D bulk. Binning the \(r_{ij}\) values into a histogram provides a visual representation of the bond length distributions (Fig. \ref{fig:pair_dist}).

The data implies that values $r_{ij}<1.2$ include the primary coordination shell and exclude the secondary coordination shells for all lattices and elements. By analyzing the vertical height of each bin in the PDF histogram, we could directly count the number of bonds associated with each lattice. For the flat lattices, the $CN$s were as expected (3 for $hc$, 4 for $sq$, and 6 for $hex$). For the buckled lattices, the $CN$ was 6 for $bhex$, 8 for $bsq$, and 9 for $bhc$ for all considered elements. Apart from $bhex$, these numbers demonstrate the profound impact of buckling on the local atomic environment, introducing additional complexity compared to their flat counterparts.

%---------------------------- Structual Properties --------------------------------------------------
\subsection{Structural properties}
\begin{figure}[p]
\centering
\includegraphics[width= \columnwidth]
{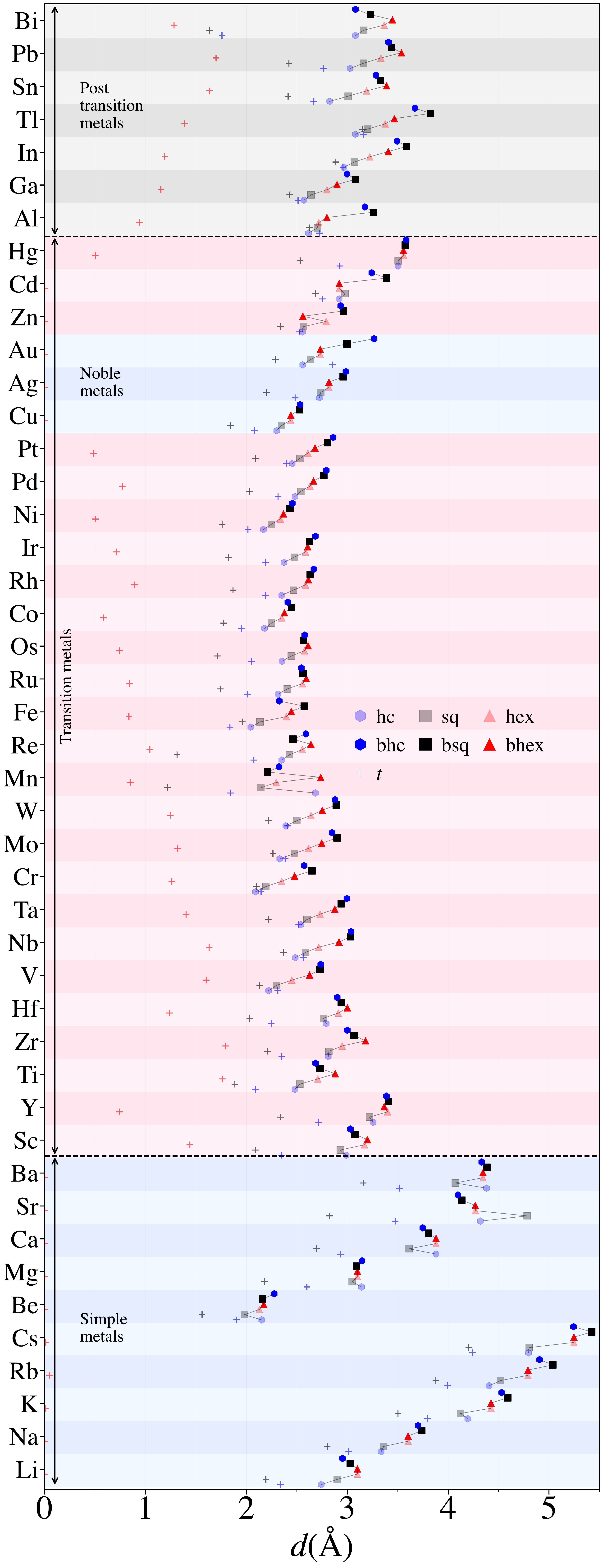}
\caption{Bond lengths ($d$) of each element and lattice. The elements are grouped for clarity. The $+$ sign indicates the buckled lattices' thickness ($t$).}
\label{fig:bond_length}
\end{figure}

\begin{figure}[p]
\centering
\includegraphics[width= \columnwidth]
{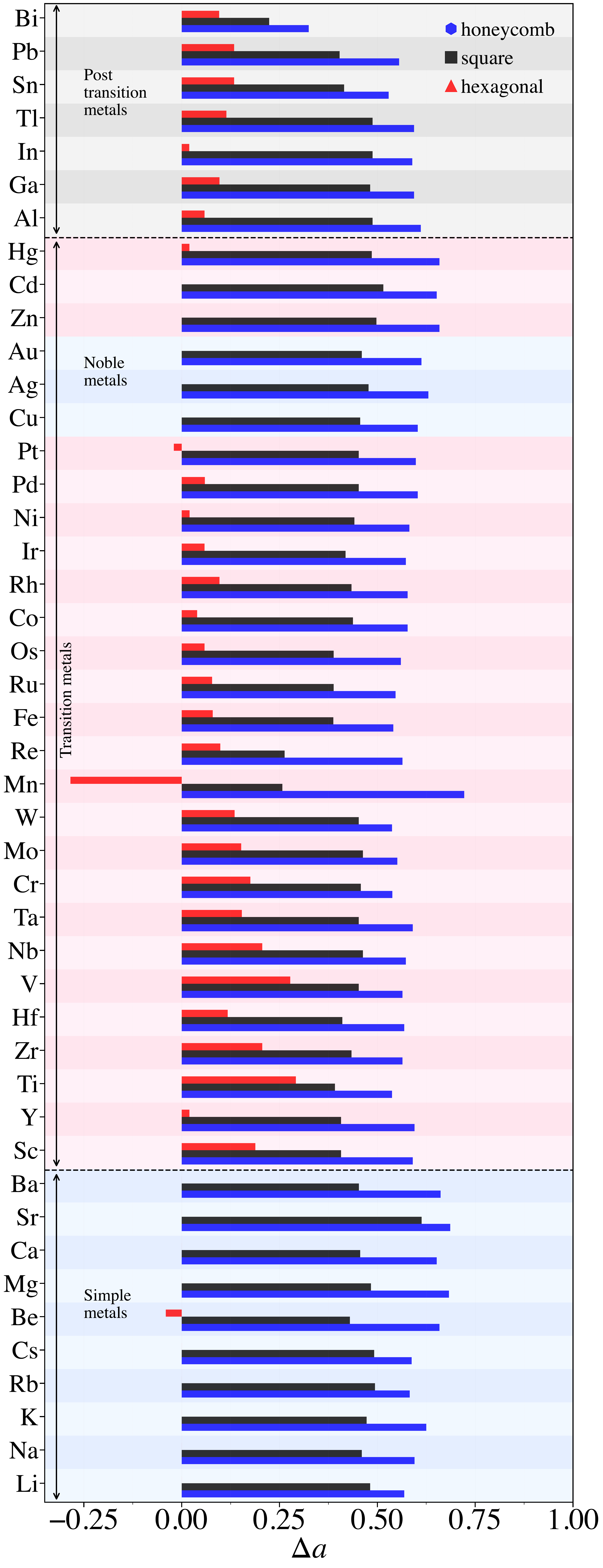}
\caption{Relative area change due to buckling \( \Delta a =(a_{f} - a_{b})/a_{f}\). A larger value means that buckling shrinks the lattice more.}
\label{fig:area_rel}
\end{figure}

We continue our discussion by focusing on the geometrical features of the metallenes, summarized in Figs. \ref{fig:bond_length} and \ref{fig:area_rel}.
%the results for simple metals, the elements from Groups 1 and 2 of the periodic table.
\paragraph{Simple metals:}
Noticeably, the structural analysis of six lattice configurations reveals the absence of out-of-plane buckling in buckled hexagonal ($bhex$) lattices. This absence can be attributed to two factors: the lack of $d$-valence electrons in simple metals and the geometrically inherent stability and frustrated symmetry of $hex$ lattices against buckling. Consequently, for simple metals, $bhex$ is equivalent to $hex$.

Regarding bond lengths in flat lattices, $hex$ lattices generally exhibit the largest bond lengths, Sr($sq$) being an exception. Group 1 elements predominantly show the shortest bonds in the $hc$ lattice, barring K, while the $sq$ lattice in group 2 elements favors the shortest bonds, except for Sr. In buckled lattices, $bsq$ typically yields larger bond lengths, except for Be and Mg.

Buckled lattices are characterized by their thickness, defined as the vertical separation between atoms layers. The $bhc$ lattice is consistently thicker than the $bsq$ lattice, with Rb and Cs exhibiting a small difference between the two lattices. This observation infers that the reduced in-plane atomic area in $bhc$ lattices promotes out-of-plane buckling as a strain-relief mechanism. In $bhc$ and $bsq$ lattices, thickness correlates directly with atomic size, which we here characterize by the covalent radius. Conversely, the increased atomic area in hexagonal geometries allows atoms to maintain a planar configuration, minimizing out-of-plane deformations.
Compared to flat lattices, the buckled lattices have smaller unit cells, and this difference in the area acquired is maximum in $bhc$, followed by $bsq$. The area per atom for $bhex$ and $hex$ are nearly identical for simple metals, except for Be, which shows $4$ $\%$ ($\approx 0.2$ $ \si{\angstrom^{2}}$) greater in-plane area in $bhex$ than $hex$ lattice.

\paragraph{Transition metals:}
Compared to simple metals, metallenes of groups 3-10 are more condensed with smaller bond lengths. The presence of $d$-electrons is strong enough to disrupt the geometrically frustrated symmetry of the $hex$ geometry against buckling, making the $bhex$ and $hex$ lattices different. Still, among buckled lattices, $bhex$ is thinnest for all metallenes. 
The $bhex$ lattices show out-of-plane buckling, with the largest thickness in Zr ($1.8$ $\si{Å}$) and the smallest in Pt ($0.48$ $\si{Å}$). The $bhc$ lattice is again the thickest, except for Fe, which is the thickest in $bsq$. The absolute difference in thickness of $bsq$ and $bhc$ was largest for Re and smallest for Cr.

Early $4d$-series metallenes have the largest bond lengths and thicknesses, followed by $3d$- and $5d$-series. Like for simple metals, the thicknesses of early transition metals correlate directly with atomic sizes. 
This trend continues until the middle of the transition series, after which it reverses: metals with larger covalent radii are thinner up to $\text{Co}$-group. Also, the $4d$-series elements (Pd and Cd) are thinner than their $5d$-series counterparts despite having larger covalent radii.
Moreover, Zr and Hf have significantly different thicknesses despite the similar covalent radii.

Transition metallenes also contract upon buckling. The largest lattice area is observed in Mn($bhc$), followed by the group 12 elements Zn($bhc$), Hg($bhc$), and Cd($bhc$). Within $bsq$ lattices, Mn has the smallest and Cd($bsq$) has the largest lattice contraction upon buckling. Ti behaves curiously: it shows the smallest contraction \( bhc \) but the largest contraction in the \( bhex \) lattice. Lattice expansion occurred only in Mn($bhex$) and Pt($bhex$). 
\paragraph{Noble metals:}
Noble metals (Cu, Ag, Au) have filled \( d\)-orbitals and properties mainly governed by $s$- and $p$-electrons$-$even at reduced dimensionality. Like simple metals and unlike other transition metals, noble metals have structurally degenerate $hex$ and $bhex$ lattices. Bond lengths increase with coordination number, while the thickness of buckled lattices, which peaks in $bhc$ lattices, positively correlates with the element's electron affinity. Buckling increases atomic densities, with Ag(\( bhc \)) showing the largest relative change to $hc$. The relative change in atomic density (\( \Delta a \)) is directly proportional to covalent radii for both \( bhc \) and \( bsq \).

\paragraph{Post-transition metals:}
For groups 13-15 metallenes, bond lengths increase with coordination number in flat lattices. In buckled lattices, the bond lengths of Sn, Pb, and Bi positively correlated with the atom size. As for transition metals, the $hex$ and $bhex$ lattices for post-transition metals are geometrically different. The largest thickness is for Pb($bhex$) ($1.70$ $\text{\AA}$) and smallest for Al($bhex$) ($0.94$ $\text{\AA}$). However, for all elements, the largest thickness occurs in $bhc$, followed by $bsq$ and $bhex$. For buckled lattices, the largest bond lengths for group 13 occur in $bsq$ and for groups 14 and 15 in $bhex$. The buckling of $hc$ and $sq$ lattices contracts the area the most for Al and the least for Bi. However, the buckling of $hex$ contracts the area the most for Sn and Pb, and the least for In.

\paragraph{Conclusions on structural properties:}
To summarize, the trends in the structural properties of metallenes are governed by atoms' electronic configurations and atomic sizes. The absence of outer \( d\)-electrons in simple metals and the filled $d$-shells in noble metals dictate their structural properties. The $s$- and $p$-electronic shells are unable to alter the hexagonal lattice symmetry, which leads to the inability to buckle the $hex$ lattice. On the contrary, for transition metals, the presence of partially filled \( d\)- electrons can disrupt the $hex$ lattice symmetry and cause sufficient out-of-plane deformation to make the \( hex \) and \( bhex \) lattices non-degenerate. Although lacking \( d\)-electrons, post-transition metals also show distinct \( hex \) and \( bhex \) structures, presumably due to both partially filled $p$-orbitals and larger atomic sizes. The degree of out-of-plane buckling varies, with \( bhc \) showing the largest thickness, followed by \( bsq \), and finally by \( bhex \). Thickness variation correlates with the atomic size and electron configuration, impacting lattice contraction upon buckling. Additionally, thickness correlates with atomic size and electron affinity in early transition metals and noble metals. These observations illustrate the sensitivity of lattice geometries on detailed electronic configurations.

%-------------------------------------Energetics -----------------------------------------------------%

\subsection{Energetics}
\begin{figure}[p]
\centering
\includegraphics[width= \columnwidth]{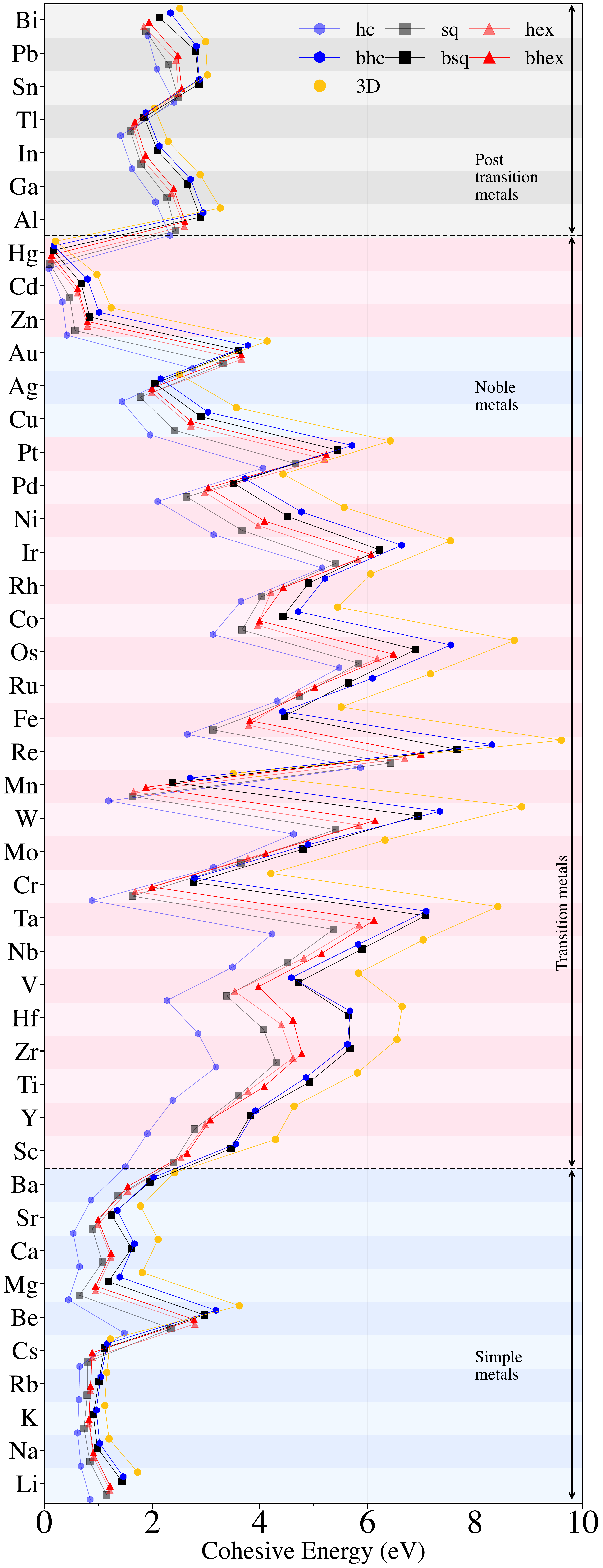}
\caption{Cohesive energies of each element and lattice.
Energies for 3D bulk are shown for comparison.}
\label{fig:coh}
\end{figure}

\begin{figure}[p]
\centering
\includegraphics[width= \columnwidth]{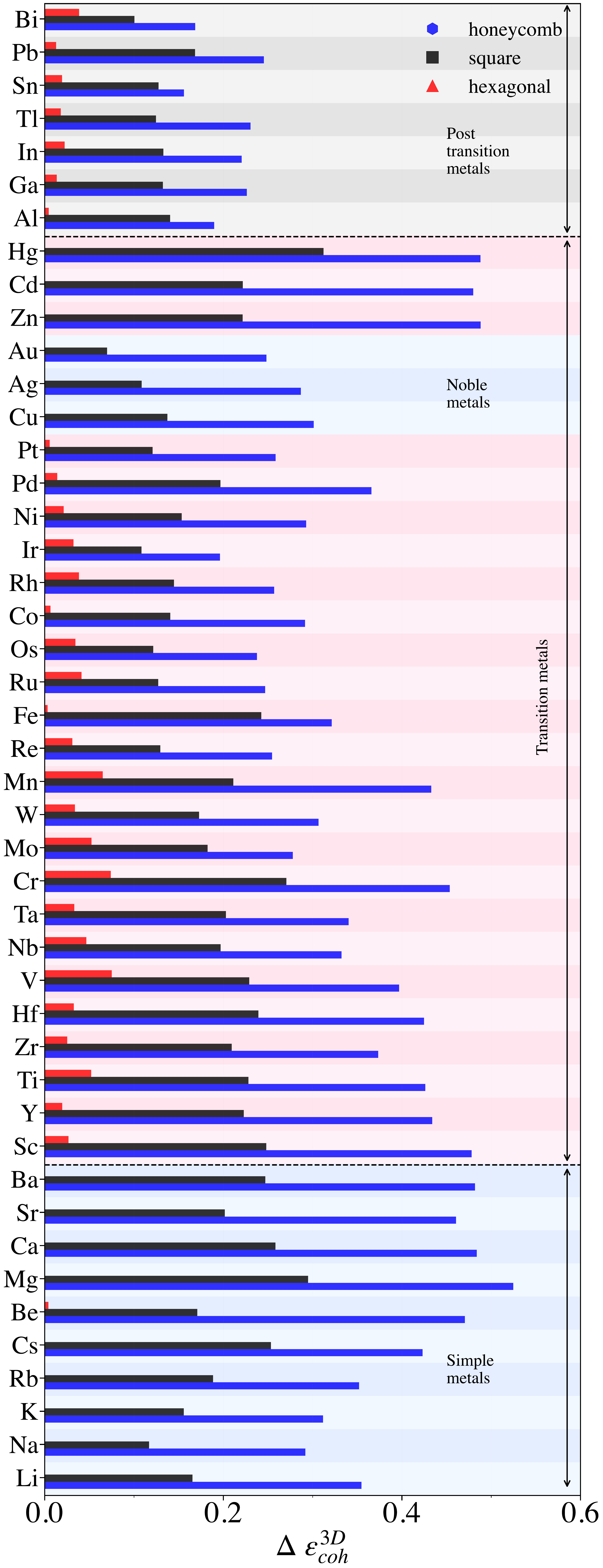}
\caption{Normalized cohesive energy differences of buckled and flat lattices \(\Delta \varepsilon_{coh}^{3D} = ({E_{coh}^{buckled} - E_{coh}^{flat}})/{E_{coh}^{3D}}\). A large \(\Delta \varepsilon_{coh}^{3D}\) means a lattice is energetically more prone to buckle.}
\label{fig:coh_relative}
\end{figure}

\begin{figure}[t!]
\centering
\includegraphics[width= \columnwidth]{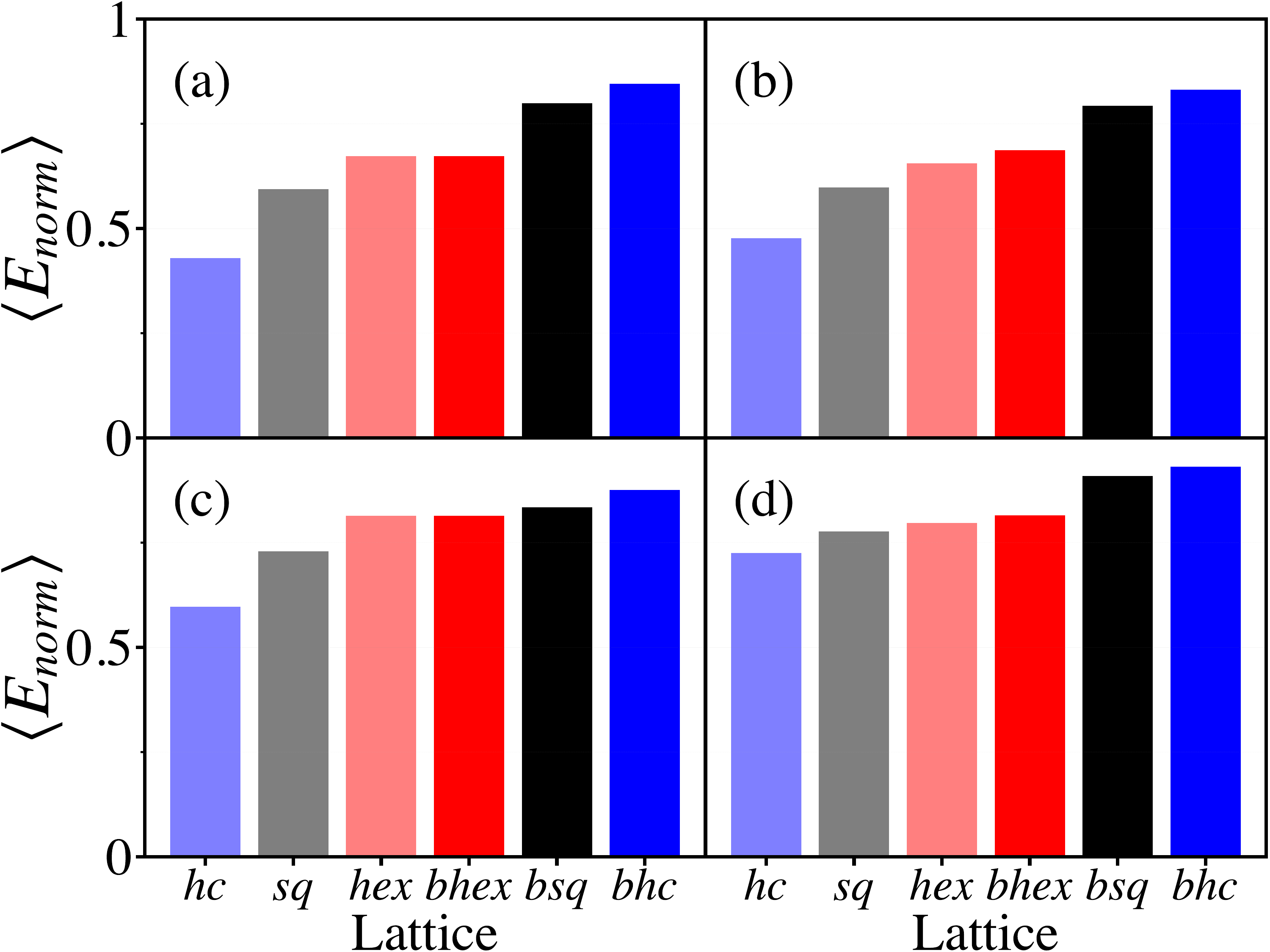}
\caption{The averages of cohesion normalized wrt. 3D bulk for (a) simple metals, (b) transition metals, (c) noble metals, and (d) post-transition metals.}
\label{fig:coh_norm}
\end{figure}

Next, we continue to investigate the energetic properties of metallenes, with results summarized in Figs. \ref{fig:coh}-\ref{fig:coh_norm}.
\paragraph{Simple metals:}
There is an evident correlation between the coordination number and cohesive energy across different lattices. With the smallest $CN=3$, the $hc$ lattice has the lowest cohesion, while the $bhc$ lattice with $CN=9$ has the highest cohesion of all lattices for all metallenes. Be($bhex$) shows a slight decrease ($0.02$ $\si{eV}$) in cohesion relative to it Be($hex$), consistent with the differences in thickness and area per atom. The transition from flat to buckled structures results in significant variations in energetic profiles, $hc$ having more pronounced effects compared to $sq$. For instance, the buckling of Mg($hc$) increases cohesion energy threefold and the buckling of Mg($sq$) increases cohesion energy by a factor of $1.8$.

Moreover, the energy change due to the buckling of K($hc$) is at least $33\%$ greater than the energy change due to the buckling of K($sq$). Such differences show how cohesion differences due to buckling are intimately tied to the lattice and the respective changes in coordination numbers. Due to its large $CN$ and geometric frustration, the buckling of $hex$ does not provide the same energetic advantage as that of $hc$ or $sq$. Therefore, smaller $CN$s make buckling more favorable in $hc$ and $sq$ lattices. 

The largest value of \(\Delta \varepsilon_{coh}^{3D}\) occurs for Mg, which implies that Mg prefers the buckled ($bhc$ and $bsq$) configurations more than any other simple metal. Na has the smallest value of \(\Delta \varepsilon_{coh}^{3D}\), suggesting that the flat lattice might also be realizable or exist in dynamic equilibrium between the two lattices as proposed in Ref.~\cite{Gentle}, depending on external factors like temperature or lateral tension.

\paragraph{Transition metals:}
Cohesive energies show similar correlations with coordination numbers as simple metals. The $hc$ lattice has the smallest cohesive energy and $bhc$ highest except for Ti, Zr, V, Nb, and Fe, which have $bsq$ ground states. The $bhex$ and $hex$ cohesion energies are different for all metallenes for groups 3-10, $bhex$ being more stable. The largest absolute cohesion energy difference between $bhex$ and $hex$ is for V ($0.44$ $\si{eV}$). The $hex$ lattice is the most energetically stable flat lattice for all metallenes, except for Ru for which $sq$ is the most stable flat lattice.

The buckling of flat lattices increases stability, with $bhc$ lattices benefiting energetically more than $bsq$ or $bhex$. For instance, the buckling increases the cohesive energy by a factor of $3$ in Cr($hc$), by a factor of $1.7$ in Cr($sq$), and by a factor of $1.2$ for Cr($hex$). The different factors show how the geometry-specific bond angles and the changes in $CN$ affect energetics; a flat lattice with a larger buckling angle benefits energetically more from buckling.

The normalized difference \(\Delta \varepsilon_{coh}^{3D}\) shows that buckling of $hc$ is most energetically stable for all metallenes. Although a few elements have $bsq$ ground state, they still exhibit a larger \(\Delta \varepsilon_{coh}^{3D}\) for the $hc$ compared to $sq$, indicating a better relative stability for $hc$. The largest $\Delta \varepsilon_{coh}^{3D}$ for elements depends on the lattice: the lattice is $hc$ for Zn, $sq$ for Cr, and $hex$ for V. However, Ir has the lowest $\Delta \varepsilon_{coh}^{3D}$ across all lattices, highlighting its distinct behavior.  
\paragraph{Noble metals:}
Cohesive energy increases with $CN$, except for Au, for whose cohesion in $hex$ is larger than $bsq$. Upon buckling the largest change in $E_{coh}$ is for Cu($bhc$) ($1.07$ $\si{eV}$) and smallest for Ag($bsq$) ($0.27$ $\si{eV}$). The low value of \(\Delta \varepsilon_{coh}^{3D}\) for Au means that buckled and flat lattices are energetically exceptionally similar and could coexist \cite{LiquidGold2D, Plenty_of_motion, Gentle}; such trend agrees with experimentally observed 2D gold, goldene \cite{taSituFabricationFreestanding2021a, kashiwayaSynthesisGoldeneComprising2024a, Sharma2022}. The largest \(\Delta \varepsilon_{coh}^{3D}\) is for Cu, indicating its tendency to buckle. 

\paragraph{Post-transition metals:}
For most elements, the cohesive energy increases with increasing $CN$, except for flat lattices of Bi, where the cohesive energy decreases with increasing $CN$. Thus, for Bi, the largest $E_{coh}$ is in $bhc$, and the smallest in $hex$. For all lattices, $E_{coh}$ is the largest for Al and the smallest for Tl. Just like the 3D bulk cohesion is nearly identical for Sn and Pb, so are the cohesion energies of their $bhc$ and $bsq$ lattices.

The value of \(\Delta \varepsilon_{coh}^{3D}\) is the lowest for Sn($bhc$), Tl($bsq$), and Al($bhex$). Conversely, it is the highest for Pb($bhc$), Pb($bsq$), and for Bi($bhex$). For a given lattice, the values of $\Delta \varepsilon_{coh}^{3D}$ are similar for Tl and Pb, implying similar intrinsic energetic stability. Therefore, because Pb metallene has been realized experimentally \cite{Grazianetti_2024}, we should expect Tl metallene to be realized as well.

\paragraph{Conclusions on energetics:}

In conclusion, cohesive energies generally increase with coordination number, making the \( bhc \) lattice the most energetically stable. Buckling in $hc$ and $sq$ geometries leads to lower energy states due to enhanced atomic density and increased coordination. This stabilization upon buckling is more pronounced in lattices that can accommodate such structural adjustments without significant strain. The buckling of $hex$ is geometrically frustrated and energetically less favorable. This behavior is emphasized for simple and noble metals, where $s$- and $p$-electrons dominate bonding interactions, preventing the buckling of $hex$ altogether (Figs. \ref{fig:coh_norm}a and \ref{fig:coh_norm}c). The additional bonding interactions introduced by \(d\)-electrons in transition metals make \(bhex\) slightly more stable than the \(hex\) (Fig. \ref{fig:coh_norm}b). Post-transition metals also show increased cohesive energies for higher coordination numbers, although the large atomic sizes for some elements introduce inconsistent energetic behaviors. Overall, the bond lengths and angles in buckled metallenes are closely coupled and make their energetic trends intimately intertwined with structural properties.

%-------------------------------------Electronic structure -----------------------------------------------------%
%----------------------------Projected Density Of States--------------------------------------------------
\subsection{Electronic structure}
We now extend our discussion to metallenes' electronic structure, analyzed in terms of projected densities of states, electron density diffusiveness, and electron localization.

\subsubsection{Projected density of states (PDOS)}
\begin{figure*}[p]
\centering
\includegraphics[width= \textwidth]{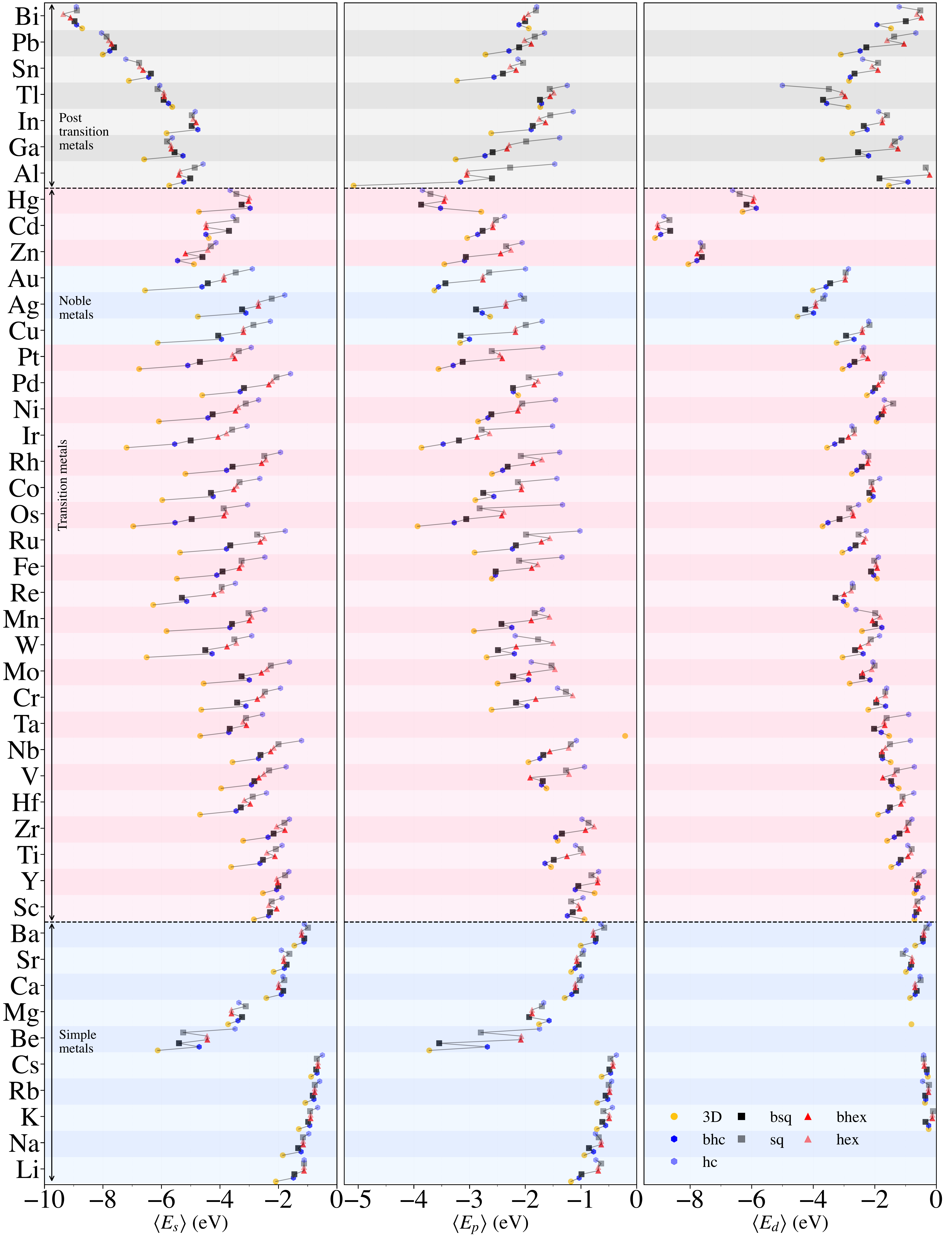}
\caption{The centers of gravity $\langle E_{l}  \rangle$ of densities of states for all elements and lattices, projected for angular momenta \( (l = s, p, d)\). The Fermi-level is set to zero. The 3D bulk is included for comparison.}
\label{fig:pdos}
\end{figure*}

We begin discussing the electronic structure properties by regarding the projected density of states; the results are summarized in Fig. \ref{fig:pdos}.

\paragraph{Simple metals:}
The $s$-orbitals of simple metals in all six lattices are occupied and deep below the Fermi-level. This stabilizing character of $s$-orbitals is particularly pronounced in alkaline earth metals Be and Mg. In contrast, Cs with its larger atomic size has $s$-states closer to the Fermi level across all lattices. The $\langle E_{s} \rangle$ of alkali metals are closest to Fermi-level in $hc$ lattice. 

Buckling often pushes $s$-states deeper for all simple metals, except for Sr($bhc$), which shifts them $0.1$~$\si{eV}$ higher in energy. The largest absolute energy shift deeper occurs for Be($bhc$) ($1.2$ $\si{eV}$) and the largest relative shift occurs for K($bhc$). The $s$-states of Cs($hc$) are closest to the Fermi level among all studied metals, while Be($bsq$) demonstrates the deepest $s$-states among simple metals.   

The $p$-orbitals lie closer to the Fermi-level than $s$-orbitals, as evident from atomic orbital ordering. This proximity suggests some role of $p$-orbitals in the directional bonding and electronic properties of atomically thin metallenes. The relative difference \((\langle E_{s}\rangle-\langle E_{p}  \rangle)/\langle E_{s}  \rangle \) was the smallest in Na($hc$) ($0.24$) and the largest in Mg($bhc$) ($0.53$).

A closer look at the $p$-orbitals in buckled lattices ($bhc$ and $bsq$) shows how in-plane $p$-states $p_{x}$ and $p_{y}$ are deeper than out-of-plane $p$-state $p_{z}$, except for Li($bhc$), Be($bhc$), Ca($bhc$), and Sr($bhc$). For flat lattices, $p_{z}$-states remain unoccupied except for Be($sq$/$hex$). As expected, the $d$-states are even closer to the Fermi level than the $p$-states, except for Sr($hc$ and $sq$) and Cs($hc$). The largest relative shift \((\langle E_{p}  \rangle - \langle E_{d}  \rangle)/\langle E_{p}  \rangle \) was observed for K($bhex$/$hex$) ($0.73$). Upon buckling, the $d$-states becomes more shallow, except for Ba($bhc$ and $bsq$), Ca($bhc$ and $bsq$), K($bsq$), and Rb($bsq$). 

For a closer look of $d$-states, we calculated the densities of states of sub-projections $t_{2g}$ ($d_{xy}$, $d_{yz}$, and $d_{xz}$) and $e_{g}$ ($d_{z^{2}}$ and $d_{x^{2}-y^{2}}$). The lattice symmetries in $hc$, $sq$, $bsq$, and $bhc$ gave degenerate $d_{xz}$ and $d_{yz}$-states; $d_{xy}$ and $d_{x^{2}-y^{2}}$ are degenerate also for $hc$ and $bhc$. We found that the deepest states are $d_{z^2}$ in $bhc$ and $bsq$ lattices, and the shallowest states are $d_{xz}$ in $bhc$, and $d_{xy}$ in $bsq$, except for Rb($bsq$) for which $d_{x^{2}-y^{2}}$ is more shallow. The $d_{z^{2}}$-state is unoccupied in $hc$, $sq$, and $hex$ lattices except for Cs($hc$, $sq$ and $hex$), Ca($hc$), and Ba($hc$ and $hex$).

We analyzed the in- and out-of-plane orbitals by calculating $\langle E_{xy} \rangle$  and $\langle E_{z} \rangle$. The $xy$-states were occupied for all metallenes, but the $z$-states of metallenes are occupied in buckled ($bsq$ and $bhc$) lattices and remain unoccupied in flat lattices, except for Cs, Be, and Ba. The $z$-states are shallower than $xy$-states in buckled lattices except for Li($bhc$), Be($bhc$), and Sr($bhc$). Buckling deepens the $xy$-states, except for Na($bhc$) and Sr($bsq$).

\paragraph{Transition metals:}
The $hc$ lattice has the shallowest $s$-states for all metallenes, except for Cd($sq$) and Hg($bhc$). The $bhc$ lattice has the deepest $s$-states, except for Cr, Mo, W, Re, and Hg, where the $s$-states are the deepest in the $bsq$ lattice. In the $3d$-series, as the number of $d$-electrons increases, the $s$-state deepen, reaching maximum for Zn. The same trend does not persist for the $4d$-series for groups 3-10. In the $5d$-series, for all lattices, Hf has the shallowest $s$-state, and Re has the deepest $s$-state, except for Ir($bhc$) and Hg($hc$). Buckling makes $s$-states deeper for all metallenes except for Hg($bhc$ and $bsq$) and Sc, Ti, Y, Zr, Hf, Ta, and Pt in the $bhex$ lattices. The largest absolute buckling-induced shifts in $\langle E_{s} \rangle$ occur for Os($bhc$), Ir($bsq$), and Zn($bhex$). 
 
The $p$-states remain unoccupied for Hf, Ta, and Re. The occupied $p$-states are shallower than $s$-states except for Mo($hc$) and Hg in all lattices. Buckling pushes $p$-states deeper except for Hg($bhc$) and Sc($bsq$ and $bhex$). The $\langle E_{p_{z}}\rangle$ is shallower than $\langle E_{p_{x}/{p_{y}}} \rangle$ for all metallenes except for Sc($bsq$), Y($bsq$), Mo($bhc$/$bsq$), Mn($hc$), Fe($bhc$), Ru($bhc$, $bsq$, $bhex$, and $hex$), Co($bhc$ and $bsq$), Rh($bhc$, $bsq$, $hex$, and $hc$), Pd($bhc$, $bsq$, and $bhex$), Cd($hc$), and Hg ($bhc$ and $bsq$).

The $d$-states are the most stable for Cd, followed by Zn and Hg across all lattices. Within groups 3-10, for a given lattice, $\langle E_{d} \rangle$ deepen as we move down the group. The $hc$ lattices have the shallowest $d$-states except for Ti($sq$), Mo($sq$), Re($sq$), Rh($hex$), Ir($hex$), Pt($hex$), Zn($bsq$), Cd($bsq$), and Hg($bhc$). Buckling pushes the $d$-states deeper except for Mn($bhc$), Hg($bhc$), Cd($bsq$), Hg($bsq$), and Sc, Y, Zr, Ta, Fe, Co, Os, and Pt in $bhex$. For the $3d$-series elements, except for Zn, $\langle E_{d} \rangle$ $>$ $\langle E_{p} \rangle$ in $bhc$ and $bsq$ lattices and for Sc in all lattices. However, for Mo, Ru, Os, and Rh, $\langle E_{d} \rangle$ $<$ $\langle E_{p} \rangle$ in all lattices. The \(3d\)-, \(4d\)-, and \(5d\)-series show a trend where \(\langle E_{d} \rangle\) decreases up to specific elements as the number of $d$-electrons increases. This trend continues up to Mn($hc$) in $3d$-series, Rh($hc$) in $4d$-series, and Os($bhc$ and $sq$) in $5d$-series; for the remaining lattices, up to Co in $3d$-series, Ru in $4d$-series, and Re in $5d$-series. The trend illustrates how $\langle E_{d} \rangle$-depends on lattice structures and the $d$-orbital electron count. 

Analyzing $\langle E_{l} \rangle$ via $d$-sub-projections showed that, as a rule, $d$-sub-orbitals deepen upon buckling. As anticipated, the $z$-states are shallower than the $xy$-states for all metallenes except for Mn($bhex$ and $hex$). The buckling of \(hc\) and \(hex\) lattices generally move the \(xy\)- and \(z\)-states deeper, with notable exceptions. In the \(hc\) geometry, the \(xy\)-states become shallower for Cr, Mo, W, Mn, and Hg, while the \(z\)-states become shallower for Mn and the group 12 elements. Similarly, the \(xy\)-states for $hex$ become shallower for Sc, Mn, Fe, Os, Co, and elements of groups 4 and 10; moreover, the \(z\)-states also become shallower for Sc, Y, Fe, Co, Rh, and Pt. The buckling of $sq$ lattice move \(z\)-states deeper and \(xy\)-states shallower for most metallenes.

\paragraph{Noble metals:}
Like in alkali metals, the $s$-states are the shallowest in $hc$ lattice. For a given lattice, the $\langle E_{s} \rangle$ is the shallowest for Ag, followed by Cu and Au. 
Buckling moves $s$-states deeper, with the largest shift in $\langle E_{s} \rangle$ for Au($bhc$) ($1.7$ $\si{eV}$). However, the relative change \((\langle E_{s}\rangle ^{buckled} - \langle E_{s}\rangle ^{flat})/\langle E_{s}\rangle ^{flat} \) was the largest for Ag($bhc$) ($0.75$) and the smallest for Au($bsq$) ($0.28$). The $p$-states are shallower than $s$-states except for Ag($hc$). Largest absolute difference between $\langle E_{s} \rangle$ and $\langle E_{p} \rangle$ is $1.1$~$\si{eV}$ for Au($bhex$ and $hex$). Buckling deepens also $p$-states, with the largest energy shift for Cu($bhc$) ($1.3$ $\si{eV}$) and the smallest for Ag($bhc$) ($0.7$ $\si{eV}$). Somewhat surprising, the \(p_z\)-states are deeper than in-plane $p_{x}/p_{y}$-states. Apart from \(p_z\)-state of Ag, all $p$-orbitals become deeper upon buckling. 

The $d$-states are deeper than $p$-states, except for Cu($bhc$ and $bsq$). Thus, like in 3D bulk, electronic interactions of noble metallenes are strongly characterized by $s$- and $p$-states. Buckling pushes \(d\)-states deeper; the shift is the largest for Cu ($bsq$) ($0.76$ $\si{eV}$) and the smallest for Ag ($bhc$) ($0.37$ $\si{eV}$). The $d$-orbital sub-projections for all lattices are deepest for Ag and shallowest for Cu. Buckling lowers the energy of $d$-projections, except for $d_{xy}$-state of Au($bsq$). Also, the $\langle E_{z} \rangle$ is shallower than $\langle E_{xy} \rangle$, and $xy$ and $z$-states become deeper upon buckling.

\paragraph{Post-transition metals:}
The $s$-states are the deepest in post-transition metallenes, with average energies around $4.5$ $\si{eV}$ below Fermi-level.
Among post-transition metallenes, for a given lattice, $s$-states are the deepest for Bi and the shallowest for In, except for Al($hc$ and $sq$). Buckling raises $s$-states, except for Al($bsq$) and Bi($bsq$). As expected, the $p$-states are again shallower than $s$-states. The absolute difference \(\langle E_{p} \rangle - \langle E_{s}\rangle\) is the largest for Bi($hex$) ($7.4$~eV) and the smallest for Al($bhc$) ($2.1$~eV). The $p_z$-states are higher than in-plane $p_{x/y}$-states for all lattices. The difference between $\langle E_{x/y} \rangle$ and $\langle E_{z} \rangle$ is the highest for Al($hex$) ($3.0$~eV) and the lowest for Bi($bhc$) ($0.3$~eV). The $p$-states deepen upon buckling, except for In, Sn, Pb in the $bhex$ lattice.

The $d$-states are deeper than $p$-states for In, Tl, Sn($bhc$, $bsq$, and $hc$), and Pb($bhc$ and $bsq$). The $d$-states remain unoccupied for Al($sq$ and $hc$). The $d$-states deepen upon buckling, except for Tl. The occupancy of $d$-sub-orbitals varies without a clear trend.

%Regarding $d$-orbital projections, 
The $xy$-states are deeper than $z$-states for all post-transition metals. Buckling deepens the $z$-states for all metallenes. Buckling of $hex$ raises the energies of $xy$-states, except for Bi. The $xy$-states are more stable in $bhc$ and $bsq$ compared to their flat counterparts, except for Sn($bhc$), Al($bsq$), and Bi($bsq$). 

\paragraph{Conclusions on the projected densities of states (PDOS):}
The PDOS analysis shows that buckling generally leads to the deepening of electronic states. This deepening is due to increased orbital overlap, stronger bonding, and results in larger cohesive energies.

For simple and noble metals, the \(s\)-orbitals lie deep below the Fermi-level while \(p\)-orbitals are shallower, indicating the role of $p$-orbitals as frontier orbitals governing much of metallenes' electronic properties. The empty or filled and deep-lying $d$-orbitals play a smaller role in the electronic structure, especially regarding external perturbations.

Partially filled $d$-orbitals make metallenes' electronic structures more complex. The trend of deepening \( d \)-orbital energies with increasing electron count up to a certain element demonstrates the correlations between electron occupancies and lattice structure across the \(3d\)-~, \(4d\)-, and \(5d\)-series. Regarding $p$-orbital projections, as expected, the \(p_z\)-orbitals are usually less stable than the in-plane \( p_{x}/p_{y} \)-orbitals, with energies highly sensitive to buckling.

%-------------------------------------Electron density thickness-----------------------------------------------------%

\subsubsection{Electron density thickness profile (FWHM)}
\begin{figure}[p]
\centering
\includegraphics[width= \columnwidth]{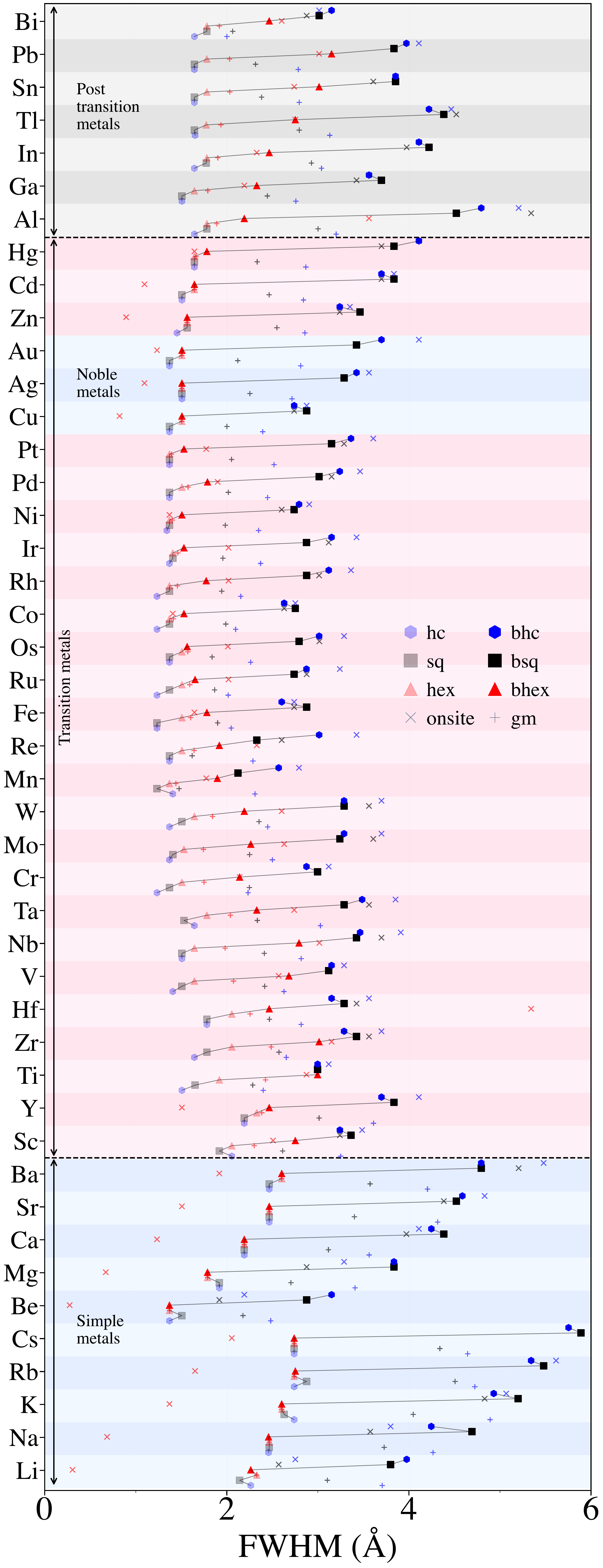}
\caption{Full width at half maximum (FWHM) of the valence electron densities along lines parallel to $z$-axis and passing through the bond centers. The $+$ sign indicates the FWHM considering only the geometrical (gm) effect on the flat lattice, and the $\times$ indicates the onsite FWHM of buckled lattices (lines passing through atoms).}
\label{fig:fwhm}
\end{figure}

To further investigate the electronic structure, from a point of view complementary to the PDOS, we analyzed the electron densities thicknesses. The results of this analysis are summarized in Fig. \ref{fig:fwhm}. 
%------------------------------SIMPLE METALS_FWHM------------------------------------------------------------
\paragraph{Simple metals:}

Among flat geometries, the $hc$ lattice shows strong in-plane electron confinement, suggesting a covalent nature of bonding (supported later by ELF analysis). In contrast, the $bsq$ lattice is consistently the thickest, except for the first elements in each group (Li and Be). Overall, the vertical diffuseness depends strongly on the lattice structure and buckling significantly readjusts the electronic cloud. FWHM plausibly increases with the atomic size, Cs($bsq$) with its largest covalent radius having the thickest electronic cloud. The relative change \( (\text{FWHM}_{bhc} - \text{FWHM}_{hc})/ \text{FWHM}_{hc}\) upon buckling is the largest for Be  ($1.3$) and smallest for Na ($0.73$). The detailed response to buckling is element- and geometry-dependent; for instance, K has a thicker electronic cloud in $hc$ compared to $sq$, but the ordering inverts upon buckling. Moreover, the geometric thickness $t$ in K($bhc$) is larger than in K($bsq$), whereas for FWHM the ordering unexpectedly is the other way. 

There are also other intricacies. FWHM for Li($sq$) is the thinnest, although Be has the smallest covalent radius. Therefore, mere atomic sizes do not fully govern electron density distributions. Be's consistently small FWHM values suggest the possibility of its planar realization, as suggested also by recent DFT studies exploring Be confinement in cycloparaphenylene pores  \cite{sangolkar2023electronic}. The onsite FWHM for buckled lattices ($bhc$ and $bsq$) shows a size-dependent trend. For smaller atoms (Li and Be), it falls between the FWHM values of flat and buckled lattices. However, it exceeds the buckled FWHM as the atomic size increases.

%------------------------------TRANSITION METALS_FWHM------------------------------------------------------------
\paragraph{Transition metals:}
For groups 3-10, the largest FWHMs are for Y in all lattices except for Zr($bhex$). Among flat lattices, $hex$ has the thickest electronic cloud, except for Mn whose cloud is the thickest for $hc$. The $3d$-series metallenes have the thickest electronic clouds in $bsq$ lattice except for V, Mn, and Ni, which are the thickest in $bhc$. The largest relative buckling-induced thickness changes \([\text{(FWHM}_{buckled} - \text{FWHM}_{flat})/\text{FWHM}_{flat}]\) were $1.5$ for Rh($hc$), $1.54$ for Cd($sq$), and $0.70$ for Nb($hex$). For Sc, $\text{FWHM}_{hc}>\text{FWHM}_{sq}$ but $\text{FWHM}_{bsq}>\text{FWHM}_{bhc}$ despite $t_{bhc}>t_{bsq}$, which demonstrates how unpredictably electronic cloud changes depending on the element. 
To understand how the physical thickness $t$ affects the electronic density thickness at the bonding region, we calculated the difference \( \text{FWHM}_{buckled} - \text{FWHM}_{gm}\). For most elements the difference is positive, meaning that most often the vertical electron diffuseness cannot be explained by geometry alone; buckling increases the vertical diffuseness more profoundly. Only Sc(\(bhc\)) and Os(\(bhex\)) have negative differences. FWHM for Fe and Cr, which have been stabilized in graphene pores experimentally \cite{2DCr, 2dironfree}, remains small for all lattices. Overall, there appears no clear correlation between FWHM and physical parameters ($t$ or atomic size). The lack of correlation suggests that electronic configurations and element-specific bonding interactions delicately determine the out-of-plane thickness of the electronic cloud. 

%------------------------------NOBLE METALS_FWHM------------------------------------------------------------

\paragraph{Noble  metals:}
The out-of-plane electronic cloud is the thinnest in Au($sq$) and the thickest in Au($bhc$). The electronic cloud thickness is almost the same for all flat lattices. This similarity in FWHM identifies that atom sizes have little effect to the out-of-plane electron distribution. Although buckling thickens the electronic cloud, the absolute difference in thickness of buckled and flat lattices is the largest for Au($bhc$) ($2.3$ $\text{\AA}$) and the smallest for Cu($bhc$) ($1.4$ $\text{\AA}$). 

%------------------------------POST TRANSITION METALS_FWHM------------------------------------------------------------

\paragraph{Post-transition metals:}

Among the same flat lattices, the electronic thicknesses are approximately equal. The electronic clouds are the thickest for Al, Pb, and Bi in $bhc$ lattice, for others in $bsq$ lattice. Among buckled lattices, the largest FWHM is for Al($bhc$ and $bsq$) and the smallest for Bi($bhc$ and $bsq$).

The FWHM correlates positively with $t$. Compared to other post-transition elements, Bi showed a modest response to buckling and is the only element that expands less than $100\%$ in the \(bhc\) lattice. The extent of non-geometrical contribution (\(\text{FWHM}-\text{FWHM}_{gm}\)) is the largest in $bsq$ lattice except for Al and Bi, where the contribution is largest in $bhc$. In the $bhc$ lattice the electronic cloud thickness onsite is larger than at the bond center, except for Bi. On contrary, in $bsq$ and $bhex$ lattices, the electronic cloud thickness onsite is smaller than at the bond center, except for Tl($bsq$), Al($bsq$ and $bhex$), and Bi($bhex$).

%------------------------------Conclusions on FWHM------------------------------------------------------------

\paragraph{Conclusions on electron density thickness:}
In summary, the electron density thickness (FWHM) of metallenes is influenced by lattice geometry, atomic size, and electronic configuration. As a rule, larger $t$ and atomic sizes imply a more diffuse electronic cloud. Still, there are many exceptions to this rule, emphasizing the large role of purely electronic factors over simple geometrical factors.

For simple metals and noble metals, the situation is simple and buckling of $hc$ and $sq$ increases FWHM dramatically. However, transition metals show no clear correlation between FWHM and geometrical factors like $t$ and atomic size. This irregularity is due to partially filled $d$-orbitals, which was also reflected in PDOS analyses. Post-transition metals showed assorted trends, with a positive correlation between FWHM and physical thickness for buckled lattices 

Overall, the analysis reveals that, while buckling and structural properties naturally influence electron density thickness, the details of \(s\)-, \(p\)-, and \(d\)-state energetics and occupations are equally important. 

%------------------------------Electron localization function (ELF)-----------------------------------------------------%
\subsubsection{Electron localization function (ELF):}
\begin{figure}[p]
\centering
\includegraphics[width=\columnwidth]{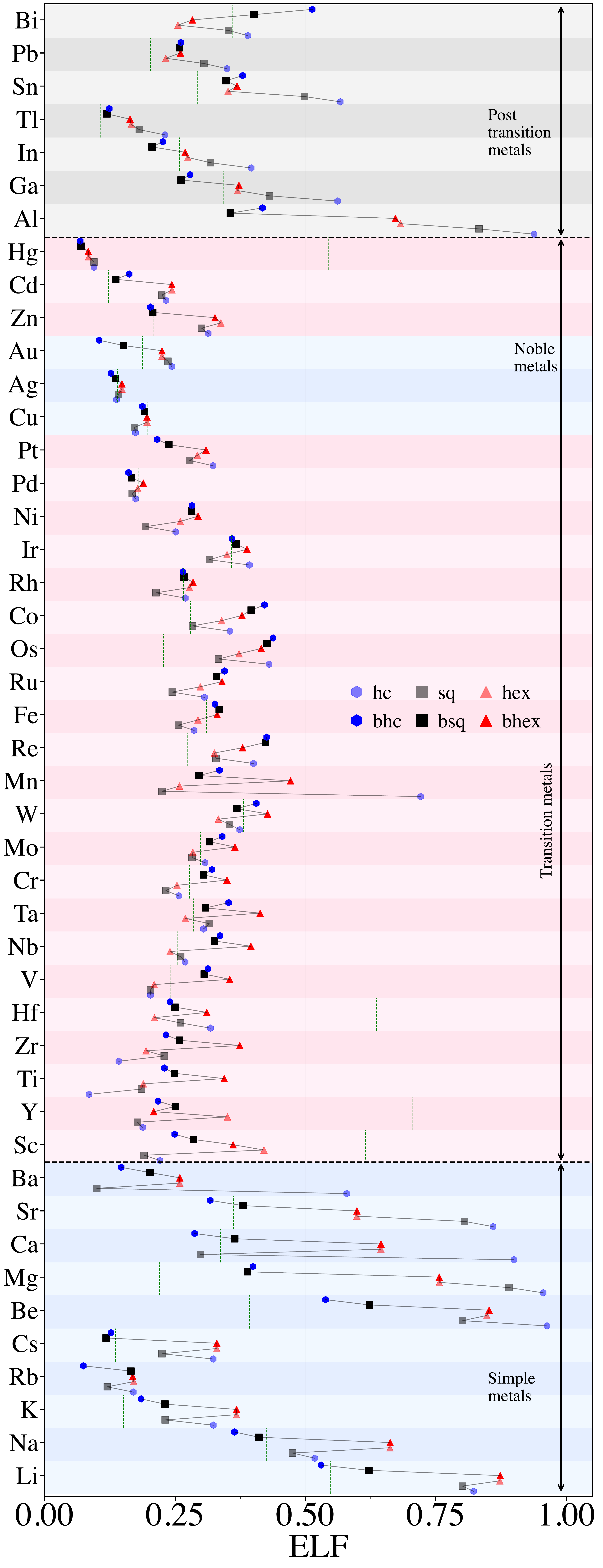}
\caption{Electron localization function (ELF) at the bond centers for all elements and lattices. The dashed green line is the ELF value at the bond center in 3D bulk.}
\label{fig:elf}
\end{figure}

As the final piece of our electronic structure analysis, we performed the ELF analysis. The ELF results are summarized in Fig.~\ref{fig:elf}. 
%------------------------------SIMPLE METALS_ELF------------------------------------------------------------
\paragraph{Simple metals:}
Recent investigations have suggested that metallic bonding may involve local orbitals not centered on the atoms but on interstitial sites instead \cite{sunChemicalInteractionsThat2023}.
The electronic localization function (ELF) here aligns with such notions for metallenes, and for most cases, local ELF at the bond center is higher for those lattices which have larger interstitial voids. The interstitial voids are the smallest for $hex$ and $bhex$ lattices, characterized by their triangular geometry, so these lattices have a greater degree of electron localization at the bond center for alkali metals but not for alkaline earth metals. This difference behaviour of alkali metals and alkaline earth metals was also noted in \cite{sunChemicalInteractionsThat2023}. Among alkaline earth metals the localization in $hc$ is striking. Here, the high ELF values at the bond center suggest small local Pauli repulsion compared to uniform electron gas, whose ELF value equals exactly $0.5$. Except for Li($bhc$ and $bsq$), buckled metallenes have ELF below $0.5$, implying strong electron delocalization. Delocalization in Rb is the strongest, followed by Ba. The $bhc$ lattice,
with its bond centers as centers of tetrahedral interstitial sites, shows stronger Pauli repulsion and greater delocalization than the flat $hex$ lattice.

Compared to 3D bulk, Li($bhc$), Na($bhc$ and $bsq$), Cs($bhc$ and $bsq$), Ca($bhc$), and Sr($bhc$) have stronger Pauli repulsion at the bond center. Further, the symmetric ELF values observed for elements and lattices, meaning equal ELFs for spin up and down electrons, confirm their non-magnetic nature in 2D as well.

%------------------------------TRANSITION METALS_ELF------------------------------------------------------------
\paragraph{Transition metals:}
All transition metallenes have a stronger Pauli repulsion at the bond center than a uniform electron gas of the same density (ELF$<0.5$). 
In groups 3-10, the ELF value across all lattices is the smallest for Pd and the largest for Os, except for W($bhex$ and $sq$). ELF is the largest for Os($bhc$) ($0.44$) and the smallest for Pd ($bhc$) ($0.16$). Within each transition series, the ELF is the largest for Sc($hc$) in $3d$-series, Nb($bhex$) in $4d$-series, Os($bhc$) in $5d$-series. The buckling of $hc$  strengthens Pauli repulsion and ELF decreases for early transition metals (Sc, Y, Ti, Zr, and Hf), group 12 metals, Ir, Pd, and Pt. The buckling of $sq$ decreases ELF for Hf, Ta, Fe, Ni, Pt, Zn, Cd, and Hg but does not affect Pd. Buckling of $hex$ geometry weakens the Pauli repulsion and ELF increases, except for Fe, Co, Ni, and Zn.

In general, the ELFs are smaller than in 3D bulk for early transition metals---electrons in metallenes are \emph{more delocalized}. In middle and late transition metallenes, ELFs depend on element and lattice type but are close to 3D bulk values. Particularly the ELF in Pd is nearly constant in all morphologies. Hg has very low ELF compared to 3D bulk, at least at this level of theory (with its highly correlated electronic structure, Hg is known to be challenging for DFT).

The ELF values were spin-dependent for many metallenes. Spin-rependence was visible in \(3d\)-series for Fe, Co, Ni(except \( sq \)), Sc($hex$ and $hc$), Ti(\(hc\)), and Mn($hbc$, $hex$, and $hc$); in $4d$-series for Y($bhc$ and $hc$), Zr(\( hc \)), Ru($bhex$ and $hex$), and Rh(\( hc \)); and in \(5d\)-series for Hf(\( hc \)), Os(\( hex \)), and Ir(\(hex\)). Thus, these metallenes may exhibit non-zero magnetic moments, as earlier predicted by Ren \emph{et al.} for flat lattices \cite{renMagnetismElementalTwodimensional2021}. For groups 4, 5, and 12 the ELF is not spin-dependent.

%------------------------------NOBLE METALS_ELF-----------------------------------------------------------

\paragraph{Noble metals:}
Like in 3D bulk, the ELFs for all lattices are small and electrons highly delocalized; the largest ELF is for Au($hc$) ($0.24$), and the smallest for Au($bhc$) ($0.10$). The ELFs for the others lie between these two values, which demonstrate a narrow spectrum of electronic delocalization; electrons in all noble metallenes are delocalized similarly. Only flat lattices of Au and Ag(except $hc$) have ELF higher than 3D bulk. 

%------------------------------POST TRANSITION METALS_ELF------------------------------------------------------------
\paragraph{Post-transition metals:}
ELFs decrease strongly as we move down the Al and Sn groups for a given lattice. The Pauli repulsion boosts and delocalization increases as the coordination number increases in flat lattices. Buckling enhances delocalization, except for Bi in all lattices and Ga, Sn, and Pb in $bhex$. Compared to 3D bulk, the electrons of Tl, Sn, and Pb are more localized in all lattices; Al, Ga, and In are more delocalized in $bhc$ and $bsq$ lattices, and Bi is more delocalized in $bhex$, $hex$, and $sq$ lattices. Electrons in the $bsq$ lattice are the most delocalized, except for Pb and Bi, where the electrons are the most delocalized in the $hex$ lattice.

%------------------------------Conclusions on ELF------------------------------------------------------------

\paragraph{Conclusions on electron localization function:}
Summarizing, buckling makes electrons more delocalized in simple, noble, and post-transition metallenes, except for Cu and Bi. This trend gets reversed for transition metals, except for Sc-, Pd-, and Zn-group metals. Most flat lattices of alkaline earth metals and Al exhibit stronger electron localization, hinting towards the most covalent-type bonding. In general, electrons in simple metals are more localized and electrons in transition metals are less localized compared to other metallenes.

%------------------------------Summary and Conclusion------------------------------------------------------------

\section{\label{sec:level5}Summary and Conclusion}

To summarize, we have investigated the electronic and structural properties of 45 elemental metallenes in six different lattices using DFT simulations. The analysis of local atomic environment gave well-defined coordination numbers for all lattices. The $hex$ and $bhex$ lattices of simple and noble metals were found to be the same. As a rule, the cohesion energies increase monotonously as a function of coordination number. Consequently, most metallenes have $bhc$ ground state lattice and $hc$ as the least stable lattice. Buckling increases stability by shifting electronic states deeper below the Fermi level. The metallenes with the highest coordination numbers also have the thickest electronic clouds. Still, detailed analysis shows that the electronic cloud thickness cannot be explained by geometrical effects alone; much of the electronic diffuseness is due to detailed shell occupation-dependent orbital contributions. The electron cloud thickness correlates with electron localization, with buckling leading to greater thickness and stronger delocalization. The spin-dependent ELF suggests the possible magnetic character of transition metallenes. In general, most structural and electronic properties are governed by systematic trends---which frequently are confirmed by one or several exceptions.

%We used 3D bulk data for comparison.

These results provide a simple and systematic yet fundamental understanding of the properties of metallenes and their connection to atomic properties such as sizes and electronic configurations. A better understanding of these properties improves not only our ability to tailor their structures by external perturbations such as electric and magnetic fields but also helps to improve their stabilities and subsequent utilization in biomedical, nanoelectronic, catalytic, and spintronic applications \cite{taSituFabricationFreestanding2021a, MetalleneBiomedical2023, MetalleneElectrocatalysisandEnergy2023, Catalysis2024}.
\vspace{5pt}
\begin{acknowledgments}
We acknowledge the Vilho, Yrjö, and Kalle Väisälä Foundation of the Finnish Academy of Science and Letters and the Jane and Aatos Erkko Foundation for funding (project EcoMet) and the Finnish Grid and Cloud Infrastructure (FGCI) and CSC — IT Center for Science for computational resources.
\end{acknowledgments}

%\bibliography{elec_str_ref}% Produces the bibliography via BibTeX.

\begin{thebibliography}{80}%
	\makeatletter
	\providecommand \@ifxundefined [1]{%
		\@ifx{#1\undefined}
	}%
	\providecommand \@ifnum [1]{%
		\ifnum #1\expandafter \@firstoftwo
		\else \expandafter \@secondoftwo
		\fi
	}%
	\providecommand \@ifx [1]{%
		\ifx #1\expandafter \@firstoftwo
		\else \expandafter \@secondoftwo
		\fi
	}%
	\providecommand \natexlab [1]{#1}%
	\providecommand \enquote  [1]{``#1''}%
	\providecommand \bibnamefont  [1]{#1}%
	\providecommand \bibfnamefont [1]{#1}%
	\providecommand \citenamefont [1]{#1}%
	\providecommand \href@noop [0]{\@secondoftwo}%
	\providecommand \href [0]{\begingroup \@sanitize@url \@href}%
	\providecommand \@href[1]{\@@startlink{#1}\@@href}%
	\providecommand \@@href[1]{\endgroup#1\@@endlink}%
	\providecommand \@sanitize@url [0]{\catcode `\\12\catcode `\$12\catcode
		`\&12\catcode `\#12\catcode `\^12\catcode `\_12\catcode `\%12\relax}%
	\providecommand \@@startlink[1]{}%
	\providecommand \@@endlink[0]{}%
	\providecommand \url  [0]{\begingroup\@sanitize@url \@url }%
	\providecommand \@url [1]{\endgroup\@href {#1}{\urlprefix }}%
	\providecommand \urlprefix  [0]{URL }%
	\providecommand \Eprint [0]{\href }%
	\providecommand \doibase [0]{https://doi.org/}%
	\providecommand \selectlanguage [0]{\@gobble}%
	\providecommand \bibinfo  [0]{\@secondoftwo}%
	\providecommand \bibfield  [0]{\@secondoftwo}%
	\providecommand \translation [1]{[#1]}%
	\providecommand \BibitemOpen [0]{}%
	\providecommand \bibitemStop [0]{}%
	\providecommand \bibitemNoStop [0]{.\EOS\space}%
	\providecommand \EOS [0]{\spacefactor3000\relax}%
	\providecommand \BibitemShut  [1]{\csname bibitem#1\endcsname}%
	\let\auto@bib@innerbib\@empty
	%</preamble>
	\bibitem [{\citenamefont {Slater}(1934)}]{slaterElectronicStructureMetals1934}%
	\BibitemOpen
	\bibfield  {author} {\bibinfo {author} {\bibfnamefont {J.~C.}\ \bibnamefont
			{Slater}},\ }\bibfield  {title} {\bibinfo {title} {The {{Electronic
					Structure}} of {{Metals}}},\ }\href
	{https://doi.org/10.1103/RevModPhys.6.209} {\bibfield  {journal} {\bibinfo
			{journal} {Reviews of Modern Physics}\ }\textbf {\bibinfo {volume} {6}},\
		\bibinfo {pages} {209} (\bibinfo {year} {1934})}\BibitemShut {NoStop}%
	\bibitem [{\citenamefont
		{Wigner}(1934)}]{wignerInteractionElectronsMetals1934}%
	\BibitemOpen
	\bibfield  {author} {\bibinfo {author} {\bibfnamefont {E.}~\bibnamefont
			{Wigner}},\ }\bibfield  {title} {\bibinfo {title} {On the {{Interaction}} of
			{{Electrons}} in {{Metals}}},\ }\href
	{https://doi.org/10.1103/PhysRev.46.1002} {\bibfield  {journal} {\bibinfo
			{journal} {Physical Review}\ }\textbf {\bibinfo {volume} {46}},\ \bibinfo
		{pages} {1002} (\bibinfo {year} {1934})}\BibitemShut {NoStop}%
	\bibitem [{\citenamefont
		{Hall}(1953)}]{hallElectronicStructureBodyCentred1953}%
	\BibitemOpen
	\bibfield  {author} {\bibinfo {author} {\bibfnamefont {G.~G.}\ \bibnamefont
			{Hall}},\ }\bibfield  {title} {\bibinfo {title} {The {{Electronic Structure}}
			of some {{Body-Centred Cubic Metals}}},\ }\href
	{https://doi.org/10.1088/0370-1298/66/12/313} {\bibfield  {journal} {\bibinfo
			{journal} {Proceedings of the Physical Society. Section A}\ }\textbf
		{\bibinfo {volume} {66}},\ \bibinfo {pages} {1162} (\bibinfo {year}
		{1953})}\BibitemShut {NoStop}%
	\bibitem [{\citenamefont
		{Pippard}(1960)}]{pippardExperimentalAnalysisElectronic1960}%
	\BibitemOpen
	\bibfield  {author} {\bibinfo {author} {\bibfnamefont {A.~B.}\ \bibnamefont
			{Pippard}},\ }\bibfield  {title} {\bibinfo {title} {Experimental analysis of
			the electronic structure of metals},\ }\href
	{https://doi.org/10.1088/0034-4885/23/1/304} {\bibfield  {journal} {\bibinfo
			{journal} {Reports on Progress in Physics}\ }\textbf {\bibinfo {volume}
			{23}},\ \bibinfo {pages} {176} (\bibinfo {year} {1960})}\BibitemShut
	{NoStop}%
	\bibitem [{\citenamefont {Lomer}(1962)}]{lomerElectronicStructureChromium1962}%
	\BibitemOpen
	\bibfield  {author} {\bibinfo {author} {\bibfnamefont {W.~M.}\ \bibnamefont
			{Lomer}},\ }\bibfield  {title} {\bibinfo {title} {Electronic {{Structure}} of
			{{Chromium Group Metals}}},\ }\href
	{https://doi.org/10.1088/0370-1328/80/2/316} {\bibfield  {journal} {\bibinfo
			{journal} {Proceedings of the Physical Society}\ }\textbf {\bibinfo {volume}
			{80}},\ \bibinfo {pages} {489} (\bibinfo {year} {1962})}\BibitemShut
	{NoStop}%
	\bibitem [{\citenamefont
		{Harrison}(1963)}]{harrisonElectronicStructureSeries1963}%
	\BibitemOpen
	\bibfield  {author} {\bibinfo {author} {\bibfnamefont {W.~A.}\ \bibnamefont
			{Harrison}},\ }\bibfield  {title} {\bibinfo {title} {Electronic {{Structure}}
			of a {{Series}} of {{Metals}}},\ }\href
	{https://doi.org/10.1103/PhysRev.131.2433} {\bibfield  {journal} {\bibinfo
			{journal} {Physical Review}\ }\textbf {\bibinfo {volume} {131}},\ \bibinfo
		{pages} {2433} (\bibinfo {year} {1963})}\BibitemShut {NoStop}%
	\bibitem [{\citenamefont {Lomer}(1969)}]{lomerElectronicStructurePure1969a}%
	\BibitemOpen
	\bibfield  {author} {\bibinfo {author} {\bibfnamefont {W.}~\bibnamefont
			{Lomer}},\ }\bibfield  {title} {\bibinfo {title} {The electronic structure of
			pure metals},\ }\href {https://doi.org/10.1016/0079-6425(69)90002-4}
	{\bibfield  {journal} {\bibinfo  {journal} {Progress in Materials Science}\
		}\textbf {\bibinfo {volume} {14}},\ \bibinfo {pages} {99} (\bibinfo {year}
		{1969})}\BibitemShut {NoStop}%
	\bibitem [{\citenamefont {Lomer}\ and\ \citenamefont
		{Gardner}(1969)}]{lomerElectronicStructurePure1969}%
	\BibitemOpen
	\bibfield  {author} {\bibinfo {author} {\bibfnamefont {W.}~\bibnamefont
			{Lomer}}\ and\ \bibinfo {author} {\bibfnamefont {W.}~\bibnamefont
			{Gardner}},\ }\bibfield  {title} {\bibinfo {title} {The electronic structure
			of pure metals},\ }\href {https://doi.org/10.1016/0079-6425(69)90004-8}
	{\bibfield  {journal} {\bibinfo  {journal} {Progress in Materials Science}\
		}\textbf {\bibinfo {volume} {14}},\ \bibinfo {pages} {141} (\bibinfo {year}
		{1969})}\BibitemShut {NoStop}%
	\bibitem [{\citenamefont {Hygh}\ and\ \citenamefont
		{Welch}(1970)}]{hyghElectronicStructureTitanium1970}%
	\BibitemOpen
	\bibfield  {author} {\bibinfo {author} {\bibfnamefont {E.~H.}\ \bibnamefont
			{Hygh}}\ and\ \bibinfo {author} {\bibfnamefont {R.~M.}\ \bibnamefont
			{Welch}},\ }\bibfield  {title} {\bibinfo {title} {Electronic {{Structure}} of
			{{Titanium}}},\ }\href {https://doi.org/10.1103/PhysRevB.1.2424} {\bibfield
		{journal} {\bibinfo  {journal} {Physical Review B}\ }\textbf {\bibinfo
			{volume} {1}},\ \bibinfo {pages} {2424} (\bibinfo {year} {1970})}\BibitemShut
	{NoStop}%
	\bibitem [{\citenamefont
		{Andersen}(1970)}]{andersenElectronicStructureFcc1970}%
	\BibitemOpen
	\bibfield  {author} {\bibinfo {author} {\bibfnamefont {O.~K.}\ \bibnamefont
			{Andersen}},\ }\bibfield  {title} {\bibinfo {title} {Electronic {{Structure}}
			of the fcc {{Transition Metals Ir}}, {{Rh}}, {{Pt}}, and {{Pd}}},\ }\href
	{https://doi.org/10.1103/PhysRevB.2.883} {\bibfield  {journal} {\bibinfo
			{journal} {Physical Review B}\ }\textbf {\bibinfo {volume} {2}},\ \bibinfo
		{pages} {883} (\bibinfo {year} {1970})}\BibitemShut {NoStop}%
	\bibitem [{\citenamefont {Louie}\ and\ \citenamefont
		{Cohen}(1974)}]{louieElectronicStructureCesium1974}%
	\BibitemOpen
	\bibfield  {author} {\bibinfo {author} {\bibfnamefont {S.~G.}\ \bibnamefont
			{Louie}}\ and\ \bibinfo {author} {\bibfnamefont {M.~L.}\ \bibnamefont
			{Cohen}},\ }\bibfield  {title} {\bibinfo {title} {Electronic structure of
			cesium under pressure},\ }\href {https://doi.org/10.1103/PhysRevB.10.3237}
	{\bibfield  {journal} {\bibinfo  {journal} {Physical Review B}\ }\textbf
		{\bibinfo {volume} {10}},\ \bibinfo {pages} {3237} (\bibinfo {year}
		{1974})}\BibitemShut {NoStop}%
	\bibitem [{\citenamefont {Woo}\ \emph {et~al.}(1975)\citenamefont {Woo},
		\citenamefont {Wang},\ and\ \citenamefont
		{Matsuura}}]{wooElectronicStructureMetals1975}%
	\BibitemOpen
	\bibfield  {author} {\bibinfo {author} {\bibfnamefont {C.~H.}\ \bibnamefont
			{Woo}}, \bibinfo {author} {\bibfnamefont {S.}~\bibnamefont {Wang}},\ and\
		\bibinfo {author} {\bibfnamefont {M.}~\bibnamefont {Matsuura}},\ }\bibfield
	{title} {\bibinfo {title} {Electronic structure of metals. {{I}}. {{Energy}}
			independent model pseudopotential formalism},\ }\href
	{https://doi.org/10.1088/0305-4608/5/10/007} {\bibfield  {journal} {\bibinfo
			{journal} {Journal of Physics F: Metal Physics}\ }\textbf {\bibinfo {volume}
			{5}},\ \bibinfo {pages} {1836} (\bibinfo {year} {1975})}\BibitemShut
	{NoStop}%
	\bibitem [{\citenamefont {Matsuura}\ \emph {et~al.}(1975)\citenamefont
		{Matsuura}, \citenamefont {Woo},\ and\ \citenamefont
		{Wang}}]{matsuuraElectronicStructureMetals1975}%
	\BibitemOpen
	\bibfield  {author} {\bibinfo {author} {\bibfnamefont {M.}~\bibnamefont
			{Matsuura}}, \bibinfo {author} {\bibfnamefont {C.~H.}\ \bibnamefont {Woo}},\
		and\ \bibinfo {author} {\bibfnamefont {S.}~\bibnamefont {Wang}},\ }\bibfield
	{title} {\bibinfo {title} {Electronic structure of metals. {{II}}. {{Phonon}}
			spectra and {{Fermi}} surface distortions},\ }\href
	{https://doi.org/10.1088/0305-4608/5/10/008} {\bibfield  {journal} {\bibinfo
			{journal} {Journal of Physics F: Metal Physics}\ }\textbf {\bibinfo {volume}
			{5}},\ \bibinfo {pages} {1849} (\bibinfo {year} {1975})}\BibitemShut
	{NoStop}%
	\bibitem [{\citenamefont {Jepsen}\ \emph {et~al.}(1975)\citenamefont {Jepsen},
		\citenamefont {Andersen},\ and\ \citenamefont
		{Mackintosh}}]{jepsenElectronicStructureHcp1975}%
	\BibitemOpen
	\bibfield  {author} {\bibinfo {author} {\bibfnamefont {O.}~\bibnamefont
			{Jepsen}}, \bibinfo {author} {\bibfnamefont {O.~K.}\ \bibnamefont
			{Andersen}},\ and\ \bibinfo {author} {\bibfnamefont {A.~R.}\ \bibnamefont
			{Mackintosh}},\ }\bibfield  {title} {\bibinfo {title} {Electronic structure
			of hcp transition metals},\ }\href {https://doi.org/10.1103/PhysRevB.12.3084}
	{\bibfield  {journal} {\bibinfo  {journal} {Physical Review B}\ }\textbf
		{\bibinfo {volume} {12}},\ \bibinfo {pages} {3084} (\bibinfo {year}
		{1975})}\BibitemShut {NoStop}%
	\bibitem [{\citenamefont {So}\ \emph {et~al.}(1977{\natexlab{a}})\citenamefont
		{So}, \citenamefont {Kuroda}, \citenamefont {Takegahara},\ and\ \citenamefont
		{Wang}}]{soElectronicStructureMetals1977}%
	\BibitemOpen
	\bibfield  {author} {\bibinfo {author} {\bibfnamefont {C.~B.}\ \bibnamefont
			{So}}, \bibinfo {author} {\bibfnamefont {Y.}~\bibnamefont {Kuroda}}, \bibinfo
		{author} {\bibfnamefont {K.}~\bibnamefont {Takegahara}},\ and\ \bibinfo
		{author} {\bibfnamefont {S.}~\bibnamefont {Wang}},\ }\bibfield  {title}
	{\bibinfo {title} {Electronic structure of metals. {{III}}. {{Properties}} of
			the lattice and of electron transport},\ }\href
	{https://doi.org/10.1088/0305-4608/7/6/006} {\bibfield  {journal} {\bibinfo
			{journal} {Journal of Physics F: Metal Physics}\ }\textbf {\bibinfo {volume}
			{7}},\ \bibinfo {pages} {885} (\bibinfo {year}
		{1977}{\natexlab{a}})}\BibitemShut {NoStop}%
	\bibitem [{\citenamefont {So}\ \emph {et~al.}(1977{\natexlab{b}})\citenamefont
		{So}, \citenamefont {Takegahara},\ and\ \citenamefont
		{Wang}}]{soElectronicStructureMetals1977a}%
	\BibitemOpen
	\bibfield  {author} {\bibinfo {author} {\bibfnamefont {C.~B.}\ \bibnamefont
			{So}}, \bibinfo {author} {\bibfnamefont {K.}~\bibnamefont {Takegahara}},\
		and\ \bibinfo {author} {\bibfnamefont {S.}~\bibnamefont {Wang}},\ }\bibfield
	{title} {\bibinfo {title} {Electronic structure of metals. {{V}}. {{Density}}
			of states effective masses of the alkali metals},\ }\href
	{https://doi.org/10.1088/0305-4608/7/8/013} {\bibfield  {journal} {\bibinfo
			{journal} {Journal of Physics F: Metal Physics}\ }\textbf {\bibinfo {volume}
			{7}},\ \bibinfo {pages} {1453} (\bibinfo {year}
		{1977}{\natexlab{b}})}\BibitemShut {NoStop}%
	\bibitem [{\citenamefont {Lai}\ \emph {et~al.}(1978)\citenamefont {Lai},
		\citenamefont {Wang},\ and\ \citenamefont
		{So}}]{laiElectronicStructureMetals1978}%
	\BibitemOpen
	\bibfield  {author} {\bibinfo {author} {\bibfnamefont {S.~K.}\ \bibnamefont
			{Lai}}, \bibinfo {author} {\bibfnamefont {S.}~\bibnamefont {Wang}},\ and\
		\bibinfo {author} {\bibfnamefont {C.~B.}\ \bibnamefont {So}},\ }\bibfield
	{title} {\bibinfo {title} {Electronic structure of metals. {{VI}}.
			{{Magnetic}} susceptibilities of simple metals},\ }\href
	{https://doi.org/10.1088/0305-4608/8/5/019} {\bibfield  {journal} {\bibinfo
			{journal} {Journal of Physics F: Metal Physics}\ }\textbf {\bibinfo {volume}
			{8}},\ \bibinfo {pages} {883} (\bibinfo {year} {1978})}\BibitemShut {NoStop}%
	\bibitem [{\citenamefont {Wang}\ and\ \citenamefont
		{So}(1979)}]{wangElectronicStructureMetals1979}%
	\BibitemOpen
	\bibfield  {author} {\bibinfo {author} {\bibfnamefont {S.}~\bibnamefont
			{Wang}}\ and\ \bibinfo {author} {\bibfnamefont {C.~B.}\ \bibnamefont {So}},\
	}\bibfield  {title} {\bibinfo {title} {Electronic structure of metals.
			{{VII}}. {{Optical}} absorption of simple metals},\ }\href
	{https://doi.org/10.1088/0305-4608/9/3/019} {\bibfield  {journal} {\bibinfo
			{journal} {Journal of Physics F: Metal Physics}\ }\textbf {\bibinfo {volume}
			{9}},\ \bibinfo {pages} {579} (\bibinfo {year} {1979})}\BibitemShut {NoStop}%
	\bibitem [{\citenamefont {Skriver}\ \emph {et~al.}(1982)\citenamefont
		{Skriver}, \citenamefont {Johansson},\ and\ \citenamefont
		{Andersen}}]{skriverCohesionElectronicStructure1982}%
	\BibitemOpen
	\bibfield  {author} {\bibinfo {author} {\bibfnamefont {H.~L.}\ \bibnamefont
			{Skriver}}, \bibinfo {author} {\bibfnamefont {B.}~\bibnamefont {Johansson}},\
		and\ \bibinfo {author} {\bibfnamefont {O.~K.}\ \bibnamefont {Andersen}},\
	}\bibfield  {title} {\bibinfo {title} {Cohesion and {{Electronic Structure}}
			of the {{Actinide Metals}}},\ }\href
	{https://doi.org/10.1088/0031-8949/1982/T1/009} {\bibfield  {journal}
		{\bibinfo  {journal} {Physica Scripta}\ }\textbf {\bibinfo {volume} {T1}},\
		\bibinfo {pages} {25} (\bibinfo {year} {1982})}\BibitemShut {NoStop}%
	\bibitem [{\citenamefont {Blaha}\ and\ \citenamefont
		{Callaway}(1985)}]{blahaElectronicStructureFermi1985}%
	\BibitemOpen
	\bibfield  {author} {\bibinfo {author} {\bibfnamefont {P.}~\bibnamefont
			{Blaha}}\ and\ \bibinfo {author} {\bibfnamefont {J.}~\bibnamefont
			{Callaway}},\ }\bibfield  {title} {\bibinfo {title} {Electronic structure and
			{{Fermi}} surface of calcium},\ }\href
	{https://doi.org/10.1103/PhysRevB.32.7664} {\bibfield  {journal} {\bibinfo
			{journal} {Physical Review B}\ }\textbf {\bibinfo {volume} {32}},\ \bibinfo
		{pages} {7664} (\bibinfo {year} {1985})}\BibitemShut {NoStop}%
	\bibitem [{\citenamefont {Tripathi}\ \emph {et~al.}(1988)\citenamefont
		{Tripathi}, \citenamefont {Brener},\ and\ \citenamefont
		{Callaway}}]{tripathiElectronicStructureRhodium1988}%
	\BibitemOpen
	\bibfield  {author} {\bibinfo {author} {\bibfnamefont {G.~S.}\ \bibnamefont
			{Tripathi}}, \bibinfo {author} {\bibfnamefont {N.~E.}\ \bibnamefont
			{Brener}},\ and\ \bibinfo {author} {\bibfnamefont {J.}~\bibnamefont
			{Callaway}},\ }\bibfield  {title} {\bibinfo {title} {Electronic structure of
			rhodium},\ }\href {https://doi.org/10.1103/PhysRevB.38.10454} {\bibfield
		{journal} {\bibinfo  {journal} {Physical Review B}\ }\textbf {\bibinfo
			{volume} {38}},\ \bibinfo {pages} {10454} (\bibinfo {year}
		{1988})}\BibitemShut {NoStop}%
	\bibitem [{\citenamefont {Blaha}\ \emph {et~al.}(1988)\citenamefont {Blaha},
		\citenamefont {Schwarz},\ and\ \citenamefont
		{Dederichs}}]{blahaElectronicStructureHcp1988}%
	\BibitemOpen
	\bibfield  {author} {\bibinfo {author} {\bibfnamefont {P.}~\bibnamefont
			{Blaha}}, \bibinfo {author} {\bibfnamefont {K.}~\bibnamefont {Schwarz}},\
		and\ \bibinfo {author} {\bibfnamefont {P.~H.}\ \bibnamefont {Dederichs}},\
	}\bibfield  {title} {\bibinfo {title} {Electronic structure of hcp metals},\
	}\href {https://doi.org/10.1103/PhysRevB.38.9368} {\bibfield  {journal}
		{\bibinfo  {journal} {Physical Review B}\ }\textbf {\bibinfo {volume} {38}},\
		\bibinfo {pages} {9368} (\bibinfo {year} {1988})}\BibitemShut {NoStop}%
	\bibitem [{\citenamefont
		{Harrison}(1989)}]{harrisonElectronicStructureProperties1989}%
	\BibitemOpen
	\bibfield  {author} {\bibinfo {author} {\bibfnamefont {W.~A.}\ \bibnamefont
			{Harrison}},\ }\href@noop {} {\emph {\bibinfo {title} {Electronic Structure
				and the Properties of Solids: The Physics of the Chemical Bond}}}\ (\bibinfo
	{publisher} {Dover Publications},\ \bibinfo {address} {New York},\ \bibinfo
	{year} {1989})\BibitemShut {NoStop}%
	\bibitem [{\citenamefont {Fuster}\ \emph {et~al.}(1990)\citenamefont {Fuster},
		\citenamefont {Tyler}, \citenamefont {Brener}, \citenamefont {Callaway},\
		and\ \citenamefont {Bagayoko}}]{fusterElectronicStructureRelated1990}%
	\BibitemOpen
	\bibfield  {author} {\bibinfo {author} {\bibfnamefont {G.}~\bibnamefont
			{Fuster}}, \bibinfo {author} {\bibfnamefont {J.~M.}\ \bibnamefont {Tyler}},
		\bibinfo {author} {\bibfnamefont {N.~E.}\ \bibnamefont {Brener}}, \bibinfo
		{author} {\bibfnamefont {J.}~\bibnamefont {Callaway}},\ and\ \bibinfo
		{author} {\bibfnamefont {D.}~\bibnamefont {Bagayoko}},\ }\bibfield  {title}
	{\bibinfo {title} {Electronic structure and related properties of silver},\
	}\href {https://doi.org/10.1103/PhysRevB.42.7322} {\bibfield  {journal}
		{\bibinfo  {journal} {Physical Review B}\ }\textbf {\bibinfo {volume} {42}},\
		\bibinfo {pages} {7322} (\bibinfo {year} {1990})}\BibitemShut {NoStop}%
	\bibitem [{\citenamefont {Lee}\ and\ \citenamefont
		{Callaway}(1993)}]{leeElectronicStructureMagnetism1993}%
	\BibitemOpen
	\bibfield  {author} {\bibinfo {author} {\bibfnamefont {K.}~\bibnamefont
			{Lee}}\ and\ \bibinfo {author} {\bibfnamefont {J.}~\bibnamefont {Callaway}},\
	}\bibfield  {title} {\bibinfo {title} {Electronic structure and magnetism of
			small {{V}} and {{Cr}} clusters},\ }\href
	{https://doi.org/10.1103/PhysRevB.48.15358} {\bibfield  {journal} {\bibinfo
			{journal} {Physical Review B}\ }\textbf {\bibinfo {volume} {48}},\ \bibinfo
		{pages} {15358} (\bibinfo {year} {1993})}\BibitemShut {NoStop}%
	\bibitem [{\citenamefont {Liu}\ and\ \citenamefont
		{Allen}(1995)}]{liuElectronicStructureSemimetals1995}%
	\BibitemOpen
	\bibfield  {author} {\bibinfo {author} {\bibfnamefont {Y.}~\bibnamefont
			{Liu}}\ and\ \bibinfo {author} {\bibfnamefont {R.~E.}\ \bibnamefont
			{Allen}},\ }\bibfield  {title} {\bibinfo {title} {Electronic structure of the
			semimetals {{Bi}} and {{Sb}}},\ }\href
	{https://doi.org/10.1103/PhysRevB.52.1566} {\bibfield  {journal} {\bibinfo
			{journal} {Physical Review B}\ }\textbf {\bibinfo {volume} {52}},\ \bibinfo
		{pages} {1566} (\bibinfo {year} {1995})}\BibitemShut {NoStop}%
	\bibitem [{\citenamefont {Xie}\ \emph {et~al.}(2001)\citenamefont {Xie},
		\citenamefont {Peng},\ and\ \citenamefont
		{Yang}}]{xieElectronicStructuresProperties2001}%
	\BibitemOpen
	\bibfield  {author} {\bibinfo {author} {\bibfnamefont {Y.-q.}\ \bibnamefont
			{Xie}}, \bibinfo {author} {\bibfnamefont {K.}~\bibnamefont {Peng}},\ and\
		\bibinfo {author} {\bibfnamefont {X.-x.}\ \bibnamefont {Yang}},\ }\bibfield
	{title} {\bibinfo {title} {Electronic structures and properties of {{Ti}},
			{{Zr}} and {{Hf}} metals},\ }\href
	{https://doi.org/10.1007/s11771-001-0031-6} {\bibfield  {journal} {\bibinfo
			{journal} {Journal of Central South University of Technology}\ }\textbf
		{\bibinfo {volume} {8}},\ \bibinfo {pages} {83} (\bibinfo {year}
		{2001})}\BibitemShut {NoStop}%
	\bibitem [{\citenamefont {Schiller}\ \emph {et~al.}(2004)\citenamefont
		{Schiller}, \citenamefont {Heber}, \citenamefont {Servedio},\ and\
		\citenamefont {Laubschat}}]{schillerElectronicStructureMg2004}%
	\BibitemOpen
	\bibfield  {author} {\bibinfo {author} {\bibfnamefont {F.}~\bibnamefont
			{Schiller}}, \bibinfo {author} {\bibfnamefont {M.}~\bibnamefont {Heber}},
		\bibinfo {author} {\bibfnamefont {V.~D.~P.}\ \bibnamefont {Servedio}},\ and\
		\bibinfo {author} {\bibfnamefont {C.}~\bibnamefont {Laubschat}},\ }\bibfield
	{title} {\bibinfo {title} {Electronic structure of {{Mg}} : {{From}}
			monolayers to bulk},\ }\href {https://doi.org/10.1103/PhysRevB.70.125106}
	{\bibfield  {journal} {\bibinfo  {journal} {Physical Review B}\ }\textbf
		{\bibinfo {volume} {70}},\ \bibinfo {pages} {125106} (\bibinfo {year}
		{2004})}\BibitemShut {NoStop}%
	\bibitem [{\citenamefont {Iota}\ \emph {et~al.}(2007)\citenamefont {Iota},
		\citenamefont {Klepeis}, \citenamefont {Yoo}, \citenamefont {Lang},
		\citenamefont {Haskel},\ and\ \citenamefont
		{Srajer}}]{iotaElectronicStructureMagnetism2007}%
	\BibitemOpen
	\bibfield  {author} {\bibinfo {author} {\bibfnamefont {V.}~\bibnamefont
			{Iota}}, \bibinfo {author} {\bibfnamefont {J.-H.~P.}\ \bibnamefont
			{Klepeis}}, \bibinfo {author} {\bibfnamefont {C.-S.}\ \bibnamefont {Yoo}},
		\bibinfo {author} {\bibfnamefont {J.}~\bibnamefont {Lang}}, \bibinfo {author}
		{\bibfnamefont {D.}~\bibnamefont {Haskel}},\ and\ \bibinfo {author}
		{\bibfnamefont {G.}~\bibnamefont {Srajer}},\ }\bibfield  {title} {\bibinfo
		{title} {Electronic structure and magnetism in compressed 3d transition
			metals},\ }\href {https://doi.org/10.1063/1.2434184} {\bibfield  {journal}
		{\bibinfo  {journal} {Applied Physics Letters}\ }\textbf {\bibinfo {volume}
			{90}},\ \bibinfo {pages} {042505} (\bibinfo {year} {2007})}\BibitemShut
	{NoStop}%
	\bibitem [{\citenamefont {Brener}\ \emph {et~al.}(2009)\citenamefont {Brener},
		\citenamefont {Callaway},\ and\ \citenamefont
		{Blaha}}]{brenerElectronicStructureSmall2009}%
	\BibitemOpen
	\bibfield  {author} {\bibinfo {author} {\bibfnamefont {N.~E.}\ \bibnamefont
			{Brener}}, \bibinfo {author} {\bibfnamefont {J.}~\bibnamefont {Callaway}},\
		and\ \bibinfo {author} {\bibfnamefont {P.}~\bibnamefont {Blaha}},\ }\bibfield
	{title} {\bibinfo {title} {Electronic structure of small sodium clusters},\
	}\href {https://doi.org/10.1002/qua.560280854} {\bibfield  {journal}
		{\bibinfo  {journal} {International Journal of Quantum Chemistry}\ }\textbf
		{\bibinfo {volume} {28}},\ \bibinfo {pages} {603} (\bibinfo {year}
		{2009})}\BibitemShut {NoStop}%
	\bibitem [{\citenamefont
		{Panda}(2012)}]{pandaElectronicStructureEquilibrium2012}%
	\BibitemOpen
	\bibfield  {author} {\bibinfo {author} {\bibfnamefont {B.~P.}\ \bibnamefont
			{Panda}},\ }\bibfield  {title} {\bibinfo {title} {Electronic structure and
			equilibrium properties of hcp titanium and zirconium},\ }\href
	{https://doi.org/10.1007/s12043-012-0301-x} {\bibfield  {journal} {\bibinfo
			{journal} {Pramana}\ }\textbf {\bibinfo {volume} {79}},\ \bibinfo {pages}
		{327} (\bibinfo {year} {2012})}\BibitemShut {NoStop}%
	\bibitem [{\citenamefont {K{\"o}nig}\ \emph {et~al.}(2021)\citenamefont
		{K{\"o}nig}, \citenamefont {Greer},\ and\ \citenamefont
		{Fahy}}]{konigElectronicPropertiesBismuth2021}%
	\BibitemOpen
	\bibfield  {author} {\bibinfo {author} {\bibfnamefont {C.}~\bibnamefont
			{K{\"o}nig}}, \bibinfo {author} {\bibfnamefont {J.~C.}\ \bibnamefont
			{Greer}},\ and\ \bibinfo {author} {\bibfnamefont {S.}~\bibnamefont {Fahy}},\
	}\bibfield  {title} {\bibinfo {title} {Electronic properties of bismuth
			nanostructures},\ }\href {https://doi.org/10.1103/PhysRevB.104.045432}
	{\bibfield  {journal} {\bibinfo  {journal} {Physical Review B}\ }\textbf
		{\bibinfo {volume} {104}},\ \bibinfo {pages} {045432} (\bibinfo {year}
		{2021})}\BibitemShut {NoStop}%
	\bibitem [{\citenamefont {Perdew}\ \emph {et~al.}(1991)\citenamefont {Perdew},
		\citenamefont {Wang},\ and\ \citenamefont {Engel}}]{perdewLiquiddrop1991}%
	\BibitemOpen
	\bibfield  {author} {\bibinfo {author} {\bibfnamefont {J.~P.}\ \bibnamefont
			{Perdew}}, \bibinfo {author} {\bibfnamefont {Y.}~\bibnamefont {Wang}},\ and\
		\bibinfo {author} {\bibfnamefont {E.}~\bibnamefont {Engel}},\ }\bibfield
	{title} {\bibinfo {title} {Liquid-drop model for crystalline metals:
			Vacancy-formation, cohesive, and face-dependent surface energies},\ }\href
	{https://doi.org/10.1103/PhysRevLett.66.508} {\bibfield  {journal} {\bibinfo
			{journal} {Phys. Rev. Lett.}\ }\textbf {\bibinfo {volume} {66}},\ \bibinfo
		{pages} {508} (\bibinfo {year} {1991})}\BibitemShut {NoStop}%
	\bibitem [{\citenamefont {Tran}\ and\ \citenamefont
		{Perdew}(2003)}]{perdewjellium2003}%
	\BibitemOpen
	\bibfield  {author} {\bibinfo {author} {\bibfnamefont {H.~T.}\ \bibnamefont
			{Tran}}\ and\ \bibinfo {author} {\bibfnamefont {J.~P.}\ \bibnamefont
			{Perdew}},\ }\bibfield  {title} {\bibinfo {title} {How metals bind: {{The}}
			deformable-jellium model with correlated electrons},\ }\href
	{https://doi.org/10.1119/1.1590653} {\bibfield  {journal} {\bibinfo
			{journal} {American Journal of Physics}\ }\textbf {\bibinfo {volume} {71}},\
		\bibinfo {pages} {1048} (\bibinfo {year} {2003})}\BibitemShut {NoStop}%
	\bibitem [{\citenamefont {Wang}\ \emph {et~al.}(2020)\citenamefont {Wang},
		\citenamefont {Park}, \citenamefont {Yu}, \citenamefont {Zhang},\ and\
		\citenamefont {Yang}}]{wangStabilitySynthesis2D2020b}%
	\BibitemOpen
	\bibfield  {author} {\bibinfo {author} {\bibfnamefont {T.}~\bibnamefont
			{Wang}}, \bibinfo {author} {\bibfnamefont {M.}~\bibnamefont {Park}}, \bibinfo
		{author} {\bibfnamefont {Q.}~\bibnamefont {Yu}}, \bibinfo {author}
		{\bibfnamefont {J.}~\bibnamefont {Zhang}},\ and\ \bibinfo {author}
		{\bibfnamefont {Y.}~\bibnamefont {Yang}},\ }\bibfield  {title} {\bibinfo
		{title} {Stability and synthesis of {{2D}} metals and alloys: A review},\
	}\href {https://doi.org/10.1016/j.mtadv.2020.100092} {\bibfield  {journal}
		{\bibinfo  {journal} {Materials Today Advances}\ }\textbf {\bibinfo {volume}
			{8}},\ \bibinfo {pages} {100092} (\bibinfo {year} {2020})}\BibitemShut
	{NoStop}%
	\bibitem [{\citenamefont {Ta}\ \emph {et~al.}(2021)\citenamefont {Ta},
		\citenamefont {Mendes}, \citenamefont {Liu}, \citenamefont {Yang},
		\citenamefont {Luo}, \citenamefont {Bachmatiuk}, \citenamefont {Gemming},
		\citenamefont {Zeng}, \citenamefont {Fu}, \citenamefont {Liu},\ and\
		\citenamefont {R{\"u}mmeli}}]{taSituFabricationFreestanding2021a}%
	\BibitemOpen
	\bibfield  {author} {\bibinfo {author} {\bibfnamefont {H.~Q.}\ \bibnamefont
			{Ta}}, \bibinfo {author} {\bibfnamefont {R.~G.}\ \bibnamefont {Mendes}},
		\bibinfo {author} {\bibfnamefont {Y.}~\bibnamefont {Liu}}, \bibinfo {author}
		{\bibfnamefont {X.}~\bibnamefont {Yang}}, \bibinfo {author} {\bibfnamefont
			{J.}~\bibnamefont {Luo}}, \bibinfo {author} {\bibfnamefont {A.}~\bibnamefont
			{Bachmatiuk}}, \bibinfo {author} {\bibfnamefont {T.}~\bibnamefont {Gemming}},
		\bibinfo {author} {\bibfnamefont {M.}~\bibnamefont {Zeng}}, \bibinfo {author}
		{\bibfnamefont {L.}~\bibnamefont {Fu}}, \bibinfo {author} {\bibfnamefont
			{L.}~\bibnamefont {Liu}},\ and\ \bibinfo {author} {\bibfnamefont {M.~H.}\
			\bibnamefont {R{\"u}mmeli}},\ }\bibfield  {title} {\bibinfo {title} {In
			{{Situ Fabrication}} of {{Freestanding Single}}-{{Atom}}-{{Thick 2D
					Metal}}/{{Metallene}} and {{2D Metal}}/ {{Metallene Oxide Membranes}}:
			{{Recent Developments}}},\ }\href {https://doi.org/10.1002/advs.202100619}
	{\bibfield  {journal} {\bibinfo  {journal} {Advanced Science}\ }\textbf
		{\bibinfo {volume} {8}},\ \bibinfo {pages} {2100619} (\bibinfo {year}
		{2021})}\BibitemShut {NoStop}%
	\bibitem [{\citenamefont {Yu}\ \emph {et~al.}(2023)\citenamefont {Yu},
		\citenamefont {Zhang},\ and\ \citenamefont
		{Yang}}]{yuTwoDimensionalMetalNanostructures2023}%
	\BibitemOpen
	\bibfield  {author} {\bibinfo {author} {\bibfnamefont {S.}~\bibnamefont
			{Yu}}, \bibinfo {author} {\bibfnamefont {C.}~\bibnamefont {Zhang}},\ and\
		\bibinfo {author} {\bibfnamefont {H.}~\bibnamefont {Yang}},\ }\bibfield
	{title} {\bibinfo {title} {Two-{{Dimensional Metal Nanostructures}}: {{From
					Theoretical Understanding}} to {{Experiment}}},\ }\href
	{https://doi.org/10.1021/acs.chemrev.2c00469} {\bibfield  {journal} {\bibinfo
			{journal} {Chemical Reviews}\ }\textbf {\bibinfo {volume} {123}},\ \bibinfo
		{pages} {3443} (\bibinfo {year} {2023})}\BibitemShut {NoStop}%
	\bibitem [{\citenamefont {Dong}\ \emph {et~al.}(2024)\citenamefont {Dong},
		\citenamefont {Lu}, \citenamefont {Lin},\ and\ \citenamefont
		{Robinson}}]{dongAirStableLargeArea2D2024}%
	\BibitemOpen
	\bibfield  {author} {\bibinfo {author} {\bibfnamefont {C.}~\bibnamefont
			{Dong}}, \bibinfo {author} {\bibfnamefont {L.-S.}\ \bibnamefont {Lu}},
		\bibinfo {author} {\bibfnamefont {Y.-C.}\ \bibnamefont {Lin}},\ and\ \bibinfo
		{author} {\bibfnamefont {J.~A.}\ \bibnamefont {Robinson}},\ }\bibfield
	{title} {\bibinfo {title} {Air-{{Stable}}, {{Large-Area 2D Metals}} and
			{{Semiconductors}}},\ }\href
	{https://doi.org/10.1021/acsnanoscienceau.3c00047} {\bibfield  {journal}
		{\bibinfo  {journal} {ACS Nanoscience Au}\ }\textbf {\bibinfo {volume} {4}},\
		\bibinfo {pages} {115} (\bibinfo {year} {2024})}\BibitemShut {NoStop}%
	\bibitem [{\citenamefont {Kashiwaya}\ \emph {et~al.}(2024)\citenamefont
		{Kashiwaya}, \citenamefont {Shi}, \citenamefont {Lu}, \citenamefont
		{Sangiovanni}, \citenamefont {Greczynski}, \citenamefont {Magnuson},
		\citenamefont {Andersson}, \citenamefont {Rosen},\ and\ \citenamefont
		{Hultman}}]{kashiwayaSynthesisGoldeneComprising2024a}%
	\BibitemOpen
	\bibfield  {author} {\bibinfo {author} {\bibfnamefont {S.}~\bibnamefont
			{Kashiwaya}}, \bibinfo {author} {\bibfnamefont {Y.}~\bibnamefont {Shi}},
		\bibinfo {author} {\bibfnamefont {J.}~\bibnamefont {Lu}}, \bibinfo {author}
		{\bibfnamefont {D.~G.}\ \bibnamefont {Sangiovanni}}, \bibinfo {author}
		{\bibfnamefont {G.}~\bibnamefont {Greczynski}}, \bibinfo {author}
		{\bibfnamefont {M.}~\bibnamefont {Magnuson}}, \bibinfo {author}
		{\bibfnamefont {M.}~\bibnamefont {Andersson}}, \bibinfo {author}
		{\bibfnamefont {J.}~\bibnamefont {Rosen}},\ and\ \bibinfo {author}
		{\bibfnamefont {L.}~\bibnamefont {Hultman}},\ }\bibfield  {title} {\bibinfo
		{title} {Synthesis of goldene comprising single-atom layer gold},\ }\href
	{https://doi.org/10.1038/s44160-024-00518-4} {\bibfield  {journal} {\bibinfo
			{journal} {Nature Synthesis}\ }\textbf {\bibinfo {volume} {3}},\ \bibinfo
		{pages} {744} (\bibinfo {year} {2024})}\BibitemShut {NoStop}%
	\bibitem [{\citenamefont {Mendes}\ \emph {et~al.}(2024)\citenamefont {Mendes},
		\citenamefont {Ta}, \citenamefont {Gemming}, \citenamefont {van Gog},
		\citenamefont {van Huis}, \citenamefont {Bachmatiuk},\ and\ \citenamefont
		{Rümmeli}}]{Zirconen_2024}%
	\BibitemOpen
	\bibfield  {author} {\bibinfo {author} {\bibfnamefont {R.~G.}\ \bibnamefont
			{Mendes}}, \bibinfo {author} {\bibfnamefont {H.~Q.}\ \bibnamefont {Ta}},
		\bibinfo {author} {\bibfnamefont {T.}~\bibnamefont {Gemming}}, \bibinfo
		{author} {\bibfnamefont {H.}~\bibnamefont {van Gog}}, \bibinfo {author}
		{\bibfnamefont {M.~A.}\ \bibnamefont {van Huis}}, \bibinfo {author}
		{\bibfnamefont {A.}~\bibnamefont {Bachmatiuk}},\ and\ \bibinfo {author}
		{\bibfnamefont {M.~H.}\ \bibnamefont {Rümmeli}},\ }\bibfield  {title}
	{\bibinfo {title} {In situ growth of suspended zirconene islets inside
			graphene pores},\ }\href
	{https://doi.org/https://doi.org/10.1002/adfm.202412889} {\bibfield
		{journal} {\bibinfo  {journal} {Advanced Functional Materials}\ }\textbf
		{\bibinfo {volume} {n/a}},\ \bibinfo {pages} {2412889} (\bibinfo {year}
		{2024})}\BibitemShut {NoStop}%
	\bibitem [{\citenamefont {Hohenberg}\ and\ \citenamefont
		{Kohn}(1964)}]{hohenbergInhomogeneousElectronGas1964a}%
	\BibitemOpen
	\bibfield  {author} {\bibinfo {author} {\bibfnamefont {P.}~\bibnamefont
			{Hohenberg}}\ and\ \bibinfo {author} {\bibfnamefont {W.}~\bibnamefont
			{Kohn}},\ }\bibfield  {title} {\bibinfo {title} {Inhomogeneous {{Electron
					Gas}}},\ }\href {https://doi.org/10.1103/PhysRev.136.B864} {\bibfield
		{journal} {\bibinfo  {journal} {Physical Review}\ }\textbf {\bibinfo {volume}
			{136}},\ \bibinfo {pages} {B864} (\bibinfo {year} {1964})}\BibitemShut
	{NoStop}%
	\bibitem [{\citenamefont {Kohn}\ and\ \citenamefont
		{Sham}(1965)}]{kohnSelfConsistentEquationsIncluding1965a}%
	\BibitemOpen
	\bibfield  {author} {\bibinfo {author} {\bibfnamefont {W.}~\bibnamefont
			{Kohn}}\ and\ \bibinfo {author} {\bibfnamefont {L.~J.}\ \bibnamefont
			{Sham}},\ }\bibfield  {title} {\bibinfo {title} {Self-{{Consistent Equations
					Including Exchange}} and {{Correlation Effects}}},\ }\href
	{https://doi.org/10.1103/PhysRev.140.A1133} {\bibfield  {journal} {\bibinfo
			{journal} {Physical Review}\ }\textbf {\bibinfo {volume} {140}},\ \bibinfo
		{pages} {A1133} (\bibinfo {year} {1965})}\BibitemShut {NoStop}%
	\bibitem [{\citenamefont {Kratzer}\ and\ \citenamefont
		{Neugebauer}(2019)}]{kratzerBasicsElectronicStructure2019}%
	\BibitemOpen
	\bibfield  {author} {\bibinfo {author} {\bibfnamefont {P.}~\bibnamefont
			{Kratzer}}\ and\ \bibinfo {author} {\bibfnamefont {J.}~\bibnamefont
			{Neugebauer}},\ }\bibfield  {title} {\bibinfo {title} {The {{Basics}} of
			{{Electronic Structure Theory}} for {{Periodic Systems}}},\ }\href
	{https://doi.org/10.3389/fchem.2019.00106} {\bibfield  {journal} {\bibinfo
			{journal} {Frontiers in Chemistry}\ }\textbf {\bibinfo {volume} {7}},\
		\bibinfo {pages} {106} (\bibinfo {year} {2019})}\BibitemShut {NoStop}%
	\bibitem [{\citenamefont {Sun}\ \emph {et~al.}(2023)\citenamefont {Sun},
		\citenamefont {Zhao}, \citenamefont {Pickard}, \citenamefont {Hemley},
		\citenamefont {Zheng},\ and\ \citenamefont
		{Miao}}]{sunChemicalInteractionsThat2023}%
	\BibitemOpen
	\bibfield  {author} {\bibinfo {author} {\bibfnamefont {Y.}~\bibnamefont
			{Sun}}, \bibinfo {author} {\bibfnamefont {L.}~\bibnamefont {Zhao}}, \bibinfo
		{author} {\bibfnamefont {C.~J.}\ \bibnamefont {Pickard}}, \bibinfo {author}
		{\bibfnamefont {R.~J.}\ \bibnamefont {Hemley}}, \bibinfo {author}
		{\bibfnamefont {Y.}~\bibnamefont {Zheng}},\ and\ \bibinfo {author}
		{\bibfnamefont {M.}~\bibnamefont {Miao}},\ }\bibfield  {title} {\bibinfo
		{title} {Chemical interactions that govern the structures of metals},\ }\href
	{https://doi.org/10.1073/pnas.2218405120} {\bibfield  {journal} {\bibinfo
			{journal} {Proceedings of the National Academy of Sciences}\ }\textbf
		{\bibinfo {volume} {120}},\ \bibinfo {pages} {e2218405120} (\bibinfo {year}
		{2023})}\BibitemShut {NoStop}%
	\bibitem [{\citenamefont {Mazzone}(2002)}]{mazzoneSystematicInitioStudy2002}%
	\BibitemOpen
	\bibfield  {author} {\bibinfo {author} {\bibfnamefont {A.}~\bibnamefont
			{Mazzone}},\ }\bibfield  {title} {\bibinfo {title} {Systematic ab initio
			study of the electronic properties of different pure and mixed systems formed
			by {{Cu}} and {{Ag}}},\ }\href
	{https://doi.org/10.1016/S0927-0256(02)00239-2} {\bibfield  {journal}
		{\bibinfo  {journal} {Computational Materials Science}\ }\textbf {\bibinfo
			{volume} {25}},\ \bibinfo {pages} {353} (\bibinfo {year} {2002})}\BibitemShut
	{NoStop}%
	\bibitem [{\citenamefont {Aguayo}\ \emph {et~al.}(2002)\citenamefont {Aguayo},
		\citenamefont {Murrieta},\ and\ \citenamefont
		{De~Coss}}]{aguayoElasticStabilityElectronic2002}%
	\BibitemOpen
	\bibfield  {author} {\bibinfo {author} {\bibfnamefont {A.}~\bibnamefont
			{Aguayo}}, \bibinfo {author} {\bibfnamefont {G.}~\bibnamefont {Murrieta}},\
		and\ \bibinfo {author} {\bibfnamefont {R.}~\bibnamefont {De~Coss}},\
	}\bibfield  {title} {\bibinfo {title} {Elastic stability and electronic
			structure of fcc {{Ti}}, {{Zr}}, and {{Hf}}: {{A}} first-principles study},\
	}\href {https://doi.org/10.1103/PhysRevB.65.092106} {\bibfield  {journal}
		{\bibinfo  {journal} {Physical Review B}\ }\textbf {\bibinfo {volume} {65}},\
		\bibinfo {pages} {092106} (\bibinfo {year} {2002})}\BibitemShut {NoStop}%
	\bibitem [{\citenamefont {Kamran}\ \emph {et~al.}(2009)\citenamefont {Kamran},
		\citenamefont {Chen},\ and\ \citenamefont
		{Chen}}]{kamranInitioExaminationDuctility2009}%
	\BibitemOpen
	\bibfield  {author} {\bibinfo {author} {\bibfnamefont {S.}~\bibnamefont
			{Kamran}}, \bibinfo {author} {\bibfnamefont {K.}~\bibnamefont {Chen}},\ and\
		\bibinfo {author} {\bibfnamefont {L.}~\bibnamefont {Chen}},\ }\bibfield
	{title} {\bibinfo {title} {{\emph{Ab Initio}} examination of ductility
			features of fcc metals},\ }\href {https://doi.org/10.1103/PhysRevB.79.024106}
	{\bibfield  {journal} {\bibinfo  {journal} {Physical Review B}\ }\textbf
		{\bibinfo {volume} {79}},\ \bibinfo {pages} {024106} (\bibinfo {year}
		{2009})}\BibitemShut {NoStop}%
	\bibitem [{\citenamefont {Chellathurai}\ \emph {et~al.}(2020)\citenamefont
		{Chellathurai}, \citenamefont {Gogovi},\ and\ \citenamefont
		{Papaconstantopoulos}}]{chellathuraiElectronicStructureTightbinding2020}%
	\BibitemOpen
	\bibfield  {author} {\bibinfo {author} {\bibfnamefont {M.}~\bibnamefont
			{Chellathurai}}, \bibinfo {author} {\bibfnamefont {G.~K.}\ \bibnamefont
			{Gogovi}},\ and\ \bibinfo {author} {\bibfnamefont {D.}~\bibnamefont
			{Papaconstantopoulos}},\ }\bibfield  {title} {\bibinfo {title} {Electronic
			structure and tight-binding molecular dynamics simulations for calcium and
			strontium},\ }\href {https://doi.org/10.1016/j.mtla.2020.100915} {\bibfield
		{journal} {\bibinfo  {journal} {Materialia}\ }\textbf {\bibinfo {volume}
			{14}},\ \bibinfo {pages} {100915} (\bibinfo {year} {2020})}\BibitemShut
	{NoStop}%
	\bibitem [{\citenamefont {Naumov}\ and\ \citenamefont
		{Hemley}(2015)}]{naumovOriginTransitionsMetallic2015}%
	\BibitemOpen
	\bibfield  {author} {\bibinfo {author} {\bibfnamefont {I.~I.}\ \bibnamefont
			{Naumov}}\ and\ \bibinfo {author} {\bibfnamefont {R.~J.}\ \bibnamefont
			{Hemley}},\ }\bibfield  {title} {\bibinfo {title} {Origin of {{Transitions}}
			between {{Metallic}} and {{Insulating States}} in {{Simple Metals}}},\ }\href
	{https://doi.org/10.1103/PhysRevLett.114.156403} {\bibfield  {journal}
		{\bibinfo  {journal} {Physical Review Letters}\ }\textbf {\bibinfo {volume}
			{114}},\ \bibinfo {pages} {156403} (\bibinfo {year} {2015})}\BibitemShut
	{NoStop}%
	\bibitem [{\citenamefont {Nevalaita}\ and\ \citenamefont
		{Koskinen}(2018{\natexlab{a}})}]{nevalaitaAtlasPropertiesElemental2018}%
	\BibitemOpen
	\bibfield  {author} {\bibinfo {author} {\bibfnamefont {J.}~\bibnamefont
			{Nevalaita}}\ and\ \bibinfo {author} {\bibfnamefont {P.}~\bibnamefont
			{Koskinen}},\ }\bibfield  {title} {\bibinfo {title} {Atlas for the properties
			of elemental two-dimensional metals},\ }\href
	{https://doi.org/10.1103/PhysRevB.97.035411} {\bibfield  {journal} {\bibinfo
			{journal} {Physical Review B}\ }\textbf {\bibinfo {volume} {97}},\ \bibinfo
		{pages} {035411} (\bibinfo {year} {2018}{\natexlab{a}})}\BibitemShut
	{NoStop}%
	\bibitem [{\citenamefont {Nevalaita}\ and\ \citenamefont
		{Koskinen}(2018{\natexlab{b}})}]{nevalaitaIdealTwodimensionalMetals2018}%
	\BibitemOpen
	\bibfield  {author} {\bibinfo {author} {\bibfnamefont {J.}~\bibnamefont
			{Nevalaita}}\ and\ \bibinfo {author} {\bibfnamefont {P.}~\bibnamefont
			{Koskinen}},\ }\bibfield  {title} {\bibinfo {title} {Beyond ideal
			two-dimensional metals: {{Edges}}, vacancies, and polarizabilities},\ }\href
	{https://doi.org/10.1103/PhysRevB.98.115433} {\bibfield  {journal} {\bibinfo
			{journal} {Physical Review B}\ }\textbf {\bibinfo {volume} {98}},\ \bibinfo
		{pages} {115433} (\bibinfo {year} {2018}{\natexlab{b}})}\BibitemShut
	{NoStop}%
	\bibitem [{\citenamefont {Nevalaita}\ and\ \citenamefont
		{Koskinen}(2019)}]{nevalaitaStabilityLimitsElemental2019}%
	\BibitemOpen
	\bibfield  {author} {\bibinfo {author} {\bibfnamefont {J.}~\bibnamefont
			{Nevalaita}}\ and\ \bibinfo {author} {\bibfnamefont {P.}~\bibnamefont
			{Koskinen}},\ }\bibfield  {title} {\bibinfo {title} {Stability limits of
			elemental {{2D}} metals in graphene pores},\ }\href
	{https://doi.org/10.1039/C9NR08533E} {\bibfield  {journal} {\bibinfo
			{journal} {Nanoscale}\ }\textbf {\bibinfo {volume} {11}},\ \bibinfo {pages}
		{22019} (\bibinfo {year} {2019})}\BibitemShut {NoStop}%
	\bibitem [{\citenamefont {Nevalaita}\ and\ \citenamefont
		{Koskinen}(2020)}]{nevalaitaFreestanding2DMetals2020}%
	\BibitemOpen
	\bibfield  {author} {\bibinfo {author} {\bibfnamefont {J.}~\bibnamefont
			{Nevalaita}}\ and\ \bibinfo {author} {\bibfnamefont {P.}~\bibnamefont
			{Koskinen}},\ }\bibfield  {title} {\bibinfo {title} {Free-standing {{2D}}
			metals from binary metal alloys},\ }\href {https://doi.org/10.1063/5.0010884}
	{\bibfield  {journal} {\bibinfo  {journal} {AIP Advances}\ }\textbf {\bibinfo
			{volume} {10}},\ \bibinfo {pages} {065327} (\bibinfo {year}
		{2020})}\BibitemShut {NoStop}%
	\bibitem [{\citenamefont
		{Ono}(2020)}]{onoDynamicalStabilityTwodimensional2020}%
	\BibitemOpen
	\bibfield  {author} {\bibinfo {author} {\bibfnamefont {S.}~\bibnamefont
			{Ono}},\ }\bibfield  {title} {\bibinfo {title} {Dynamical stability of
			two-dimensional metals in the periodic table},\ }\href
	{https://doi.org/10.1103/PhysRevB.102.165424} {\bibfield  {journal} {\bibinfo
			{journal} {Physical Review B}\ }\textbf {\bibinfo {volume} {102}},\ \bibinfo
		{pages} {165424} (\bibinfo {year} {2020})}\BibitemShut {NoStop}%
	\bibitem [{\citenamefont {Ono}(2021)}]{onoComprehensiveSearchBuckled2021}%
	\BibitemOpen
	\bibfield  {author} {\bibinfo {author} {\bibfnamefont {S.}~\bibnamefont
			{Ono}},\ }\bibfield  {title} {\bibinfo {title} {Comprehensive search for
			buckled honeycomb binary compounds based on noble metals ({{Cu}}, {{Ag}}, and
			{{Au}})},\ }\href {https://doi.org/10.1103/PhysRevMaterials.5.104004}
	{\bibfield  {journal} {\bibinfo  {journal} {Physical Review Materials}\
		}\textbf {\bibinfo {volume} {5}},\ \bibinfo {pages} {104004} (\bibinfo {year}
		{2021})}\BibitemShut {NoStop}%
	\bibitem [{\citenamefont {Ren}\ \emph {et~al.}(2021)\citenamefont {Ren},
		\citenamefont {Hu}, \citenamefont {Shao}, \citenamefont {Hu}, \citenamefont
		{Huang},\ and\ \citenamefont
		{Shi}}]{renMagnetismElementalTwodimensional2021}%
	\BibitemOpen
	\bibfield  {author} {\bibinfo {author} {\bibfnamefont {Y.}~\bibnamefont
			{Ren}}, \bibinfo {author} {\bibfnamefont {L.}~\bibnamefont {Hu}}, \bibinfo
		{author} {\bibfnamefont {Y.}~\bibnamefont {Shao}}, \bibinfo {author}
		{\bibfnamefont {Y.}~\bibnamefont {Hu}}, \bibinfo {author} {\bibfnamefont
			{L.}~\bibnamefont {Huang}},\ and\ \bibinfo {author} {\bibfnamefont
			{X.}~\bibnamefont {Shi}},\ }\bibfield  {title} {\bibinfo {title} {Magnetism
			of elemental two-dimensional metals},\ }\href
	{https://doi.org/10.1039/D1TC00438G} {\bibfield  {journal} {\bibinfo
			{journal} {Journal of Materials Chemistry C}\ }\textbf {\bibinfo {volume}
			{9}},\ \bibinfo {pages} {4554} (\bibinfo {year} {2021})}\BibitemShut
	{NoStop}%
	\bibitem [{\citenamefont {Sangolkar}\ \emph
		{et~al.}(2022{\natexlab{a}})\citenamefont {Sangolkar}, \citenamefont {Jha},\
		and\ \citenamefont {Pawar}}]{sangolkarDensityFunctionalTheory2022}%
	\BibitemOpen
	\bibfield  {author} {\bibinfo {author} {\bibfnamefont {A.~A.}\ \bibnamefont
			{Sangolkar}}, \bibinfo {author} {\bibfnamefont {S.}~\bibnamefont {Jha}},\
		and\ \bibinfo {author} {\bibfnamefont {R.}~\bibnamefont {Pawar}},\ }\bibfield
	{title} {\bibinfo {title} {Density {{Functional Theory}}-{{Based
					Calculations}} for {{2D Hexagonal Lanthanide Metals}}},\ }\href
	{https://doi.org/10.1002/adts.202200057} {\bibfield  {journal} {\bibinfo
			{journal} {Advanced Theory and Simulations}\ }\textbf {\bibinfo {volume}
			{5}},\ \bibinfo {pages} {2200057} (\bibinfo {year}
		{2022}{\natexlab{a}})}\BibitemShut {NoStop}%
	\bibitem [{\citenamefont {Sangolkar}\ \emph
		{et~al.}(2022{\natexlab{b}})\citenamefont {Sangolkar}, \citenamefont
		{Agrawal},\ and\ \citenamefont
		{Pawar}}]{sangolkarProspectusThicknessDependent2022}%
	\BibitemOpen
	\bibfield  {author} {\bibinfo {author} {\bibfnamefont {A.~A.}\ \bibnamefont
			{Sangolkar}}, \bibinfo {author} {\bibfnamefont {R.}~\bibnamefont {Agrawal}},\
		and\ \bibinfo {author} {\bibfnamefont {R.}~\bibnamefont {Pawar}},\ }\bibfield
	{title} {\bibinfo {title} {A prospectus for thickness dependent electronic
			properties of two-dimensional metals using density functional theory
			calculation},\ }\href {https://doi.org/10.1002/qua.26982} {\bibfield
		{journal} {\bibinfo  {journal} {International Journal of Quantum Chemistry}\
		}\textbf {\bibinfo {volume} {122}},\ \bibinfo {pages} {e26982} (\bibinfo
		{year} {2022}{\natexlab{b}})}\BibitemShut {NoStop}%
	\bibitem [{\citenamefont {Sangolkar}\ and\ \citenamefont
		{Pawar}(2022)}]{sangolkarStructureStabilityProperties2022}%
	\BibitemOpen
	\bibfield  {author} {\bibinfo {author} {\bibfnamefont {A.~A.}\ \bibnamefont
			{Sangolkar}}\ and\ \bibinfo {author} {\bibfnamefont {R.}~\bibnamefont
			{Pawar}},\ }\bibfield  {title} {\bibinfo {title} {Structure, {{Stability}},
			{{Properties}}, and {{Application}} of {{Atomically Thin Coinage Metal
					Flatland}} in {{Graphene Pore}}: {{A Density Functional Theory
					Calculation}}},\ }\href {https://doi.org/10.1002/pssb.202100489} {\bibfield
		{journal} {\bibinfo  {journal} {physica status solidi (b)}\ }\textbf
		{\bibinfo {volume} {259}},\ \bibinfo {pages} {2100489} (\bibinfo {year}
		{2022})}\BibitemShut {NoStop}%
	\bibitem [{\citenamefont {Abidi}\ and\ \citenamefont
		{Koskinen}(2023)}]{abidiElectronicStructureElasticity2023}%
	\BibitemOpen
	\bibfield  {author} {\bibinfo {author} {\bibfnamefont {K.~R.}\ \bibnamefont
			{Abidi}}\ and\ \bibinfo {author} {\bibfnamefont {P.}~\bibnamefont
			{Koskinen}},\ }\bibfield  {title} {\bibinfo {title} {Electronic structure and
			elasticity of two-dimensional metals of group 10: {{A DFT}} study},\ }\href
	{https://doi.org/10.1088/1742-6596/2518/1/012006} {\bibfield  {journal}
		{\bibinfo  {journal} {Journal of Physics: Conference Series}\ }\textbf
		{\bibinfo {volume} {2518}},\ \bibinfo {pages} {012006} (\bibinfo {year}
		{2023})}\BibitemShut {NoStop}%
	\bibitem [{\citenamefont {Abidi}\ and\ \citenamefont
		{Koskinen}(2024)}]{Gentle}%
	\BibitemOpen
	\bibfield  {author} {\bibinfo {author} {\bibfnamefont {K.~R.}\ \bibnamefont
			{Abidi}}\ and\ \bibinfo {author} {\bibfnamefont {P.}~\bibnamefont
			{Koskinen}},\ }\bibfield  {title} {\bibinfo {title} {Gentle tension
			stabilizes atomically thin metallenes},\ }\href
	{https://doi.org/10.1039/D4NR03266G} {\bibfield  {journal} {\bibinfo
			{journal} {Nanoscale}\ }\textbf {\bibinfo {volume} {16}},\ \bibinfo {pages}
		{19649} (\bibinfo {year} {2024})}\BibitemShut {NoStop}%
	\bibitem [{\citenamefont {Greeley}\ \emph {et~al.}(2002)\citenamefont
		{Greeley}, \citenamefont {N{\o}rskov},\ and\ \citenamefont
		{Mavrikakis}}]{greeleyLECTRONICTRUCTUREATALYSIS2002}%
	\BibitemOpen
	\bibfield  {author} {\bibinfo {author} {\bibfnamefont {J.}~\bibnamefont
			{Greeley}}, \bibinfo {author} {\bibfnamefont {J.~K.}\ \bibnamefont
			{N{\o}rskov}},\ and\ \bibinfo {author} {\bibfnamefont {M.}~\bibnamefont
			{Mavrikakis}},\ }\bibfield  {title} {\bibinfo {title} {Electronic structure
			and catalysis on metals surfaces},\ }\href
	{https://doi.org/10.1146/annurev.physchem.53.100301.131630} {\bibfield
		{journal} {\bibinfo  {journal} {Annual Review of Physical Chemistry}\
		}\textbf {\bibinfo {volume} {53}},\ \bibinfo {pages} {319} (\bibinfo {year}
		{2002})}\BibitemShut {NoStop}%
	\bibitem [{\citenamefont
		{Molenda}(2017)}]{molendaElectronicStructureEngineering2017}%
	\BibitemOpen
	\bibfield  {author} {\bibinfo {author} {\bibfnamefont {J.}~\bibnamefont
			{Molenda}},\ }\bibfield  {title} {\bibinfo {title} {Electronic structure
			`engineering' in the development of materials for {{Li-ion}} and {{Na-ion}}
			batteries},\ }\href {https://doi.org/10.1088/2043-6254/aa5955} {\bibfield
		{journal} {\bibinfo  {journal} {Advances in Natural Sciences: Nanoscience and
				Nanotechnology}\ }\textbf {\bibinfo {volume} {8}},\ \bibinfo {pages} {015007}
		(\bibinfo {year} {2017})}\BibitemShut {NoStop}%
	\bibitem [{\citenamefont {Xiong}\ \emph {et~al.}(2022)\citenamefont {Xiong},
		\citenamefont {Qiu}, \citenamefont {Peng}, \citenamefont {Liu},\ and\
		\citenamefont {Chu}}]{xiongElectronicStructuralEngineering2022}%
	\BibitemOpen
	\bibfield  {author} {\bibinfo {author} {\bibfnamefont {L.}~\bibnamefont
			{Xiong}}, \bibinfo {author} {\bibfnamefont {Y.}~\bibnamefont {Qiu}}, \bibinfo
		{author} {\bibfnamefont {X.}~\bibnamefont {Peng}}, \bibinfo {author}
		{\bibfnamefont {Z.}~\bibnamefont {Liu}},\ and\ \bibinfo {author}
		{\bibfnamefont {P.~K.}\ \bibnamefont {Chu}},\ }\bibfield  {title} {\bibinfo
		{title} {Electronic structural engineering of transition metal-based
			electrocatalysts for the hydrogen evolution reaction},\ }\href
	{https://doi.org/10.1016/j.nanoen.2022.107882} {\bibfield  {journal}
		{\bibinfo  {journal} {Nano Energy}\ }\textbf {\bibinfo {volume} {104}},\
		\bibinfo {pages} {107882} (\bibinfo {year} {2022})}\BibitemShut {NoStop}%
	\bibitem [{\citenamefont {Smidstrup}\ \emph {et~al.}(2020)\citenamefont
		{Smidstrup}, \citenamefont {Markussen}, \citenamefont {Vancraeyveld},
		\citenamefont {Wellendorff}, \citenamefont {Schneider}, \citenamefont
		{Gunst}, \citenamefont {Verstichel}, \citenamefont {Stradi}, \citenamefont
		{Khomyakov}, \citenamefont {{Vej-Hansen}}, \citenamefont {Lee}, \citenamefont
		{Chill}, \citenamefont {Rasmussen}, \citenamefont {Penazzi}, \citenamefont
		{Corsetti}, \citenamefont {Ojanper{\"a}}, \citenamefont {Jensen},
		\citenamefont {Palsgaard}, \citenamefont {Martinez}, \citenamefont {Blom},
		\citenamefont {Brandbyge},\ and\ \citenamefont
		{Stokbro}}]{smidstrupQuantumATKIntegratedPlatform2020}%
	\BibitemOpen
	\bibfield  {author} {\bibinfo {author} {\bibfnamefont {S.}~\bibnamefont
			{Smidstrup}}, \bibinfo {author} {\bibfnamefont {T.}~\bibnamefont
			{Markussen}}, \bibinfo {author} {\bibfnamefont {P.}~\bibnamefont
			{Vancraeyveld}}, \bibinfo {author} {\bibfnamefont {J.}~\bibnamefont
			{Wellendorff}}, \bibinfo {author} {\bibfnamefont {J.}~\bibnamefont
			{Schneider}}, \bibinfo {author} {\bibfnamefont {T.}~\bibnamefont {Gunst}},
		\bibinfo {author} {\bibfnamefont {B.}~\bibnamefont {Verstichel}}, \bibinfo
		{author} {\bibfnamefont {D.}~\bibnamefont {Stradi}}, \bibinfo {author}
		{\bibfnamefont {P.~A.}\ \bibnamefont {Khomyakov}}, \bibinfo {author}
		{\bibfnamefont {U.~G.}\ \bibnamefont {{Vej-Hansen}}}, \bibinfo {author}
		{\bibfnamefont {M.-E.}\ \bibnamefont {Lee}}, \bibinfo {author} {\bibfnamefont
			{S.~T.}\ \bibnamefont {Chill}}, \bibinfo {author} {\bibfnamefont
			{F.}~\bibnamefont {Rasmussen}}, \bibinfo {author} {\bibfnamefont
			{G.}~\bibnamefont {Penazzi}}, \bibinfo {author} {\bibfnamefont
			{F.}~\bibnamefont {Corsetti}}, \bibinfo {author} {\bibfnamefont
			{A.}~\bibnamefont {Ojanper{\"a}}}, \bibinfo {author} {\bibfnamefont
			{K.}~\bibnamefont {Jensen}}, \bibinfo {author} {\bibfnamefont {M.~L.~N.}\
			\bibnamefont {Palsgaard}}, \bibinfo {author} {\bibfnamefont {U.}~\bibnamefont
			{Martinez}}, \bibinfo {author} {\bibfnamefont {A.}~\bibnamefont {Blom}},
		\bibinfo {author} {\bibfnamefont {M.}~\bibnamefont {Brandbyge}},\ and\
		\bibinfo {author} {\bibfnamefont {K.}~\bibnamefont {Stokbro}},\ }\bibfield
	{title} {\bibinfo {title} {{{QuantumATK}}: An integrated platform of
			electronic and atomic-scale modelling tools},\ }\href
	{https://doi.org/10.1088/1361-648X/ab4007} {\bibfield  {journal} {\bibinfo
			{journal} {Journal of Physics: Condensed Matter}\ }\textbf {\bibinfo {volume}
			{32}},\ \bibinfo {pages} {015901} (\bibinfo {year} {2020})}\BibitemShut
	{NoStop}%
	\bibitem [{\citenamefont {Van~Setten}\ \emph {et~al.}(2018)\citenamefont
		{Van~Setten}, \citenamefont {Giantomassi}, \citenamefont {Bousquet},
		\citenamefont {Verstraete}, \citenamefont {Hamann}, \citenamefont {Gonze},\
		and\ \citenamefont {Rignanese}}]{vansettenPseudoDojoTrainingGrading2018}%
	\BibitemOpen
	\bibfield  {author} {\bibinfo {author} {\bibfnamefont {M.}~\bibnamefont
			{Van~Setten}}, \bibinfo {author} {\bibfnamefont {M.}~\bibnamefont
			{Giantomassi}}, \bibinfo {author} {\bibfnamefont {E.}~\bibnamefont
			{Bousquet}}, \bibinfo {author} {\bibfnamefont {M.}~\bibnamefont
			{Verstraete}}, \bibinfo {author} {\bibfnamefont {D.}~\bibnamefont {Hamann}},
		\bibinfo {author} {\bibfnamefont {X.}~\bibnamefont {Gonze}},\ and\ \bibinfo
		{author} {\bibfnamefont {G.-M.}\ \bibnamefont {Rignanese}},\ }\bibfield
	{title} {\bibinfo {title} {The {{PseudoDojo}}: {{Training}} and grading a 85
			element optimized norm-conserving pseudopotential table},\ }\href
	{https://doi.org/10.1016/j.cpc.2018.01.012} {\bibfield  {journal} {\bibinfo
			{journal} {Computer Physics Communications}\ }\textbf {\bibinfo {volume}
			{226}},\ \bibinfo {pages} {39} (\bibinfo {year} {2018})}\BibitemShut
	{NoStop}%
	\bibitem [{\citenamefont {Perdew}\ \emph {et~al.}(1996)\citenamefont {Perdew},
		\citenamefont {Burke},\ and\ \citenamefont
		{Ernzerhof}}]{perdewGeneralizedGradientApproximation1996a}%
	\BibitemOpen
	\bibfield  {author} {\bibinfo {author} {\bibfnamefont {J.~P.}\ \bibnamefont
			{Perdew}}, \bibinfo {author} {\bibfnamefont {K.}~\bibnamefont {Burke}},\ and\
		\bibinfo {author} {\bibfnamefont {M.}~\bibnamefont {Ernzerhof}},\ }\bibfield
	{title} {\bibinfo {title} {Generalized {{Gradient Approximation Made
					Simple}}},\ }\href {https://doi.org/10.1103/PhysRevLett.77.3865} {\bibfield
		{journal} {\bibinfo  {journal} {Physical Review Letters}\ }\textbf {\bibinfo
			{volume} {77}},\ \bibinfo {pages} {3865} (\bibinfo {year}
		{1996})}\BibitemShut {NoStop}%
	\bibitem [{\citenamefont {Monkhorst}\ and\ \citenamefont
		{Pack}(1976)}]{monkhorstSpecialPointsBrillouinzone1976a}%
	\BibitemOpen
	\bibfield  {author} {\bibinfo {author} {\bibfnamefont {H.~J.}\ \bibnamefont
			{Monkhorst}}\ and\ \bibinfo {author} {\bibfnamefont {J.~D.}\ \bibnamefont
			{Pack}},\ }\bibfield  {title} {\bibinfo {title} {Special points for
			{{Brillouin-zone}} integrations},\ }\href
	{https://doi.org/10.1103/PhysRevB.13.5188} {\bibfield  {journal} {\bibinfo
			{journal} {Physical Review B}\ }\textbf {\bibinfo {volume} {13}},\ \bibinfo
		{pages} {5188} (\bibinfo {year} {1976})}\BibitemShut {NoStop}%
	\bibitem [{\citenamefont {Abidi}\ and\ \citenamefont
		{Koskinen}(2022)}]{abidiOptimizingDensityfunctionalSimulations2022a}%
	\BibitemOpen
	\bibfield  {author} {\bibinfo {author} {\bibfnamefont {K.~R.}\ \bibnamefont
			{Abidi}}\ and\ \bibinfo {author} {\bibfnamefont {P.}~\bibnamefont
			{Koskinen}},\ }\bibfield  {title} {\bibinfo {title} {Optimizing
			density-functional simulations for two-dimensional metals},\ }\href
	{https://doi.org/10.1103/PhysRevMaterials.6.124004} {\bibfield  {journal}
		{\bibinfo  {journal} {Physical Review Materials}\ }\textbf {\bibinfo {volume}
			{6}},\ \bibinfo {pages} {124004} (\bibinfo {year} {2022})}\BibitemShut
	{NoStop}%
	\bibitem [{\citenamefont {Liu}\ and\ \citenamefont
		{Nocedal}(1989)}]{liuLimitedMemoryBFGS1989}%
	\BibitemOpen
	\bibfield  {author} {\bibinfo {author} {\bibfnamefont {D.~C.}\ \bibnamefont
			{Liu}}\ and\ \bibinfo {author} {\bibfnamefont {J.}~\bibnamefont {Nocedal}},\
	}\bibfield  {title} {\bibinfo {title} {On the limited memory {{BFGS}} method
			for large scale optimization},\ }\href {https://doi.org/10.1007/BF01589116}
	{\bibfield  {journal} {\bibinfo  {journal} {Mathematical Programming}\
		}\textbf {\bibinfo {volume} {45}},\ \bibinfo {pages} {503} (\bibinfo {year}
		{1989})}\BibitemShut {NoStop}%
	\bibitem [{\citenamefont {Koskinen}\ \emph {et~al.}(2007)\citenamefont
		{Koskinen}, \citenamefont {H\"akkinen}, \citenamefont {Huber}, \citenamefont
		{von Issendorff},\ and\ \citenamefont {Moseler}}]{LiquidGold2D}%
	\BibitemOpen
	\bibfield  {author} {\bibinfo {author} {\bibfnamefont {P.}~\bibnamefont
			{Koskinen}}, \bibinfo {author} {\bibfnamefont {H.}~\bibnamefont
			{H\"akkinen}}, \bibinfo {author} {\bibfnamefont {B.}~\bibnamefont {Huber}},
		\bibinfo {author} {\bibfnamefont {B.}~\bibnamefont {von Issendorff}},\ and\
		\bibinfo {author} {\bibfnamefont {M.}~\bibnamefont {Moseler}},\ }\bibfield
	{title} {\bibinfo {title} {Liquid-liquid phase coexistence in gold clusters:
			2d or not 2d?},\ }\href {https://doi.org/10.1103/PhysRevLett.98.015701}
	{\bibfield  {journal} {\bibinfo  {journal} {Phys. Rev. Lett.}\ }\textbf
		{\bibinfo {volume} {98}},\ \bibinfo {pages} {015701} (\bibinfo {year}
		{2007})}\BibitemShut {NoStop}%
	\bibitem [{\citenamefont {Koskinen}\ and\ \citenamefont
		{Korhonen}(2015)}]{Plenty_of_motion}%
	\BibitemOpen
	\bibfield  {author} {\bibinfo {author} {\bibfnamefont {P.}~\bibnamefont
			{Koskinen}}\ and\ \bibinfo {author} {\bibfnamefont {T.}~\bibnamefont
			{Korhonen}},\ }\bibfield  {title} {\bibinfo {title} {Plenty of motion at the
			bottom: atomically thin liquid gold membrane},\ }\href
	{https://doi.org/10.1039/C5NR01849H} {\bibfield  {journal} {\bibinfo
			{journal} {Nanoscale}\ }\textbf {\bibinfo {volume} {7}},\ \bibinfo {pages}
		{10140} (\bibinfo {year} {2015})}\BibitemShut {NoStop}%
	\bibitem [{\citenamefont {Sharma}\ \emph {et~al.}(2022)\citenamefont {Sharma},
		\citenamefont {Pasricha}, \citenamefont {Weston}, \citenamefont {Blanton},\
		and\ \citenamefont {Jagannathan}}]{Sharma2022}%
	\BibitemOpen
	\bibfield  {author} {\bibinfo {author} {\bibfnamefont {S.~K.}\ \bibnamefont
			{Sharma}}, \bibinfo {author} {\bibfnamefont {R.}~\bibnamefont {Pasricha}},
		\bibinfo {author} {\bibfnamefont {J.}~\bibnamefont {Weston}}, \bibinfo
		{author} {\bibfnamefont {T.}~\bibnamefont {Blanton}},\ and\ \bibinfo {author}
		{\bibfnamefont {R.}~\bibnamefont {Jagannathan}},\ }\bibfield  {title}
	{\bibinfo {title} {Synthesis of self-assembled single atomic layer gold
			crystals-goldene},\ }\href {https://doi.org/10.1021/acsami.2c19743}
	{\bibfield  {journal} {\bibinfo  {journal} {ACS Applied Materials \&
				Interfaces}\ }\textbf {\bibinfo {volume} {14}},\ \bibinfo {pages} {54992}
		(\bibinfo {year} {2022})}\BibitemShut {NoStop}%
	\bibitem [{\citenamefont {Grazianetti}\ \emph {et~al.}(2024)\citenamefont
		{Grazianetti}, \citenamefont {Molle},\ and\ \citenamefont
		{Martella}}]{Grazianetti_2024}%
	\BibitemOpen
	\bibfield  {author} {\bibinfo {author} {\bibfnamefont {C.}~\bibnamefont
			{Grazianetti}}, \bibinfo {author} {\bibfnamefont {A.}~\bibnamefont {Molle}},\
		and\ \bibinfo {author} {\bibfnamefont {C.}~\bibnamefont {Martella}},\
	}\bibfield  {title} {\bibinfo {title} {The future of xenes beyond graphene:
			challenges and perspective},\ }\href
	{https://doi.org/10.1088/2053-1583/ad77e0} {\bibfield  {journal} {\bibinfo
			{journal} {2D Materials}\ }\textbf {\bibinfo {volume} {11}},\ \bibinfo
		{pages} {042005} (\bibinfo {year} {2024})}\BibitemShut {NoStop}%
	\bibitem [{\citenamefont {Sangolkar}\ \emph {et~al.}(2023)\citenamefont
		{Sangolkar}, \citenamefont {Kadiyam}, \citenamefont {Faizan}, \citenamefont
		{Chedupaka}, \citenamefont {Mucherla},\ and\ \citenamefont
		{Pawar}}]{sangolkar2023electronic}%
	\BibitemOpen
	\bibfield  {author} {\bibinfo {author} {\bibfnamefont {A.~A.}\ \bibnamefont
			{Sangolkar}}, \bibinfo {author} {\bibfnamefont {R.~K.}\ \bibnamefont
			{Kadiyam}}, \bibinfo {author} {\bibfnamefont {M.}~\bibnamefont {Faizan}},
		\bibinfo {author} {\bibfnamefont {O.}~\bibnamefont {Chedupaka}}, \bibinfo
		{author} {\bibfnamefont {R.}~\bibnamefont {Mucherla}},\ and\ \bibinfo
		{author} {\bibfnamefont {R.}~\bibnamefont {Pawar}},\ }\href@noop {}
	{\bibfield  {journal} {\bibinfo  {journal} {Physical Chemistry Chemical
				Physics}\ }\textbf {\bibinfo {volume} {25}},\ \bibinfo {pages} {23262}
		(\bibinfo {year} {2023})}\BibitemShut {NoStop}%
	\bibitem [{\citenamefont {Ta}\ \emph {et~al.}(2020)\citenamefont {Ta},
		\citenamefont {Yang}, \citenamefont {Liu}, \citenamefont {Bachmatiuk},
		\citenamefont {Mendes}, \citenamefont {Gemming}, \citenamefont {Liu},
		\citenamefont {Liu}, \citenamefont {Tokarska}, \citenamefont {Patel} \emph
		{et~al.}}]{2DCr}%
	\BibitemOpen
	\bibfield  {author} {\bibinfo {author} {\bibfnamefont {H.~Q.}\ \bibnamefont
			{Ta}}, \bibinfo {author} {\bibfnamefont {Q.~X.}\ \bibnamefont {Yang}},
		\bibinfo {author} {\bibfnamefont {S.}~\bibnamefont {Liu}}, \bibinfo {author}
		{\bibfnamefont {A.}~\bibnamefont {Bachmatiuk}}, \bibinfo {author}
		{\bibfnamefont {R.~G.}\ \bibnamefont {Mendes}}, \bibinfo {author}
		{\bibfnamefont {T.}~\bibnamefont {Gemming}}, \bibinfo {author} {\bibfnamefont
			{Y.}~\bibnamefont {Liu}}, \bibinfo {author} {\bibfnamefont {L.}~\bibnamefont
			{Liu}}, \bibinfo {author} {\bibfnamefont {K.}~\bibnamefont {Tokarska}},
		\bibinfo {author} {\bibfnamefont {R.~B.}\ \bibnamefont {Patel}}, \emph
		{et~al.},\ }\href@noop {} {\bibfield  {journal} {\bibinfo  {journal} {Nano
				Letters}\ }\textbf {\bibinfo {volume} {20}},\ \bibinfo {pages} {4354}
		(\bibinfo {year} {2020})}\BibitemShut {NoStop}%
	\bibitem [{\citenamefont {Zhao}\ \emph {et~al.}(2014)\citenamefont {Zhao},
		\citenamefont {Deng}, \citenamefont {Bachmatiuk}, \citenamefont {Sandeep},
		\citenamefont {Popov}, \citenamefont {Eckert},\ and\ \citenamefont
		{R{\"u}mmeli}}]{2dironfree}%
	\BibitemOpen
	\bibfield  {author} {\bibinfo {author} {\bibfnamefont {J.}~\bibnamefont
			{Zhao}}, \bibinfo {author} {\bibfnamefont {Q.}~\bibnamefont {Deng}}, \bibinfo
		{author} {\bibfnamefont {A.}~\bibnamefont {Bachmatiuk}}, \bibinfo {author}
		{\bibfnamefont {G.}~\bibnamefont {Sandeep}}, \bibinfo {author} {\bibfnamefont
			{A.}~\bibnamefont {Popov}}, \bibinfo {author} {\bibfnamefont
			{J.}~\bibnamefont {Eckert}},\ and\ \bibinfo {author} {\bibfnamefont {M.~H.}\
			\bibnamefont {R{\"u}mmeli}},\ }\href@noop {} {\bibfield  {journal} {\bibinfo
			{journal} {Science}\ }\textbf {\bibinfo {volume} {343}},\ \bibinfo {pages}
		{1228} (\bibinfo {year} {2014})}\BibitemShut {NoStop}%
	\bibitem [{\citenamefont {Lu}\ \emph {et~al.}(2023)\citenamefont {Lu},
		\citenamefont {Li}, \citenamefont {Miao}, \citenamefont {Wang},\ and\
		\citenamefont {Zha}}]{MetalleneBiomedical2023}%
	\BibitemOpen
	\bibfield  {author} {\bibinfo {author} {\bibfnamefont {C.}~\bibnamefont
			{Lu}}, \bibinfo {author} {\bibfnamefont {R.}~\bibnamefont {Li}}, \bibinfo
		{author} {\bibfnamefont {Z.}~\bibnamefont {Miao}}, \bibinfo {author}
		{\bibfnamefont {F.}~\bibnamefont {Wang}},\ and\ \bibinfo {author}
		{\bibfnamefont {Z.}~\bibnamefont {Zha}},\ }\bibfield  {title} {\bibinfo
		{title} {Emerging metallenes: synthesis strategies{,} biological effects and
			biomedical applications},\ }\href {https://doi.org/10.1039/D2CS00586G}
	{\bibfield  {journal} {\bibinfo  {journal} {Chem. Soc. Rev.}\ }\textbf
		{\bibinfo {volume} {52}},\ \bibinfo {pages} {2833} (\bibinfo {year}
		{2023})}\BibitemShut {NoStop}%
	\bibitem [{\citenamefont {Xie}\ \emph {et~al.}(2023)\citenamefont {Xie},
		\citenamefont {Tang}, \citenamefont {Zhang},\ and\ \citenamefont
		{Yu}}]{MetalleneElectrocatalysisandEnergy2023}%
	\BibitemOpen
	\bibfield  {author} {\bibinfo {author} {\bibfnamefont {M.}~\bibnamefont
			{Xie}}, \bibinfo {author} {\bibfnamefont {S.}~\bibnamefont {Tang}}, \bibinfo
		{author} {\bibfnamefont {B.}~\bibnamefont {Zhang}},\ and\ \bibinfo {author}
		{\bibfnamefont {G.}~\bibnamefont {Yu}},\ }\bibfield  {title} {\bibinfo
		{title} {Metallene-related materials for electrocatalysis and energy
			conversion},\ }\href {https://doi.org/10.1039/D2MH01213H} {\bibfield
		{journal} {\bibinfo  {journal} {Mater. Horiz.}\ }\textbf {\bibinfo {volume}
			{10}},\ \bibinfo {pages} {407} (\bibinfo {year} {2023})}\BibitemShut
	{NoStop}%
	\bibitem [{\citenamefont {Shahzad}\ \emph {et~al.}(2024)\citenamefont
		{Shahzad}, \citenamefont {Saeed}, \citenamefont {Rabbee}, \citenamefont
		{Marwani}, \citenamefont {Al-Humaidi}, \citenamefont {Altaf}, \citenamefont
		{Althomali}, \citenamefont {Baek}, \citenamefont {Awual},\ and\ \citenamefont
		{Rahman}}]{Catalysis2024}%
	\BibitemOpen
	\bibfield  {author} {\bibinfo {author} {\bibfnamefont {U.}~\bibnamefont
			{Shahzad}}, \bibinfo {author} {\bibfnamefont {M.}~\bibnamefont {Saeed}},
		\bibinfo {author} {\bibfnamefont {M.~F.}\ \bibnamefont {Rabbee}}, \bibinfo
		{author} {\bibfnamefont {H.~M.}\ \bibnamefont {Marwani}}, \bibinfo {author}
		{\bibfnamefont {J.~Y.}\ \bibnamefont {Al-Humaidi}}, \bibinfo {author}
		{\bibfnamefont {M.}~\bibnamefont {Altaf}}, \bibinfo {author} {\bibfnamefont
			{R.~H.}\ \bibnamefont {Althomali}}, \bibinfo {author} {\bibfnamefont {K.-H.}\
			\bibnamefont {Baek}}, \bibinfo {author} {\bibfnamefont {M.~R.}\ \bibnamefont
			{Awual}},\ and\ \bibinfo {author} {\bibfnamefont {M.~M.}\ \bibnamefont
			{Rahman}},\ }\bibfield  {title} {\bibinfo {title} {Recent progress in
			two-dimensional metallenes and their potential application as
			electrocatalyst},\ }\href
	{https://doi.org/https://doi.org/10.1016/j.jechem.2024.02.068} {\bibfield
		{journal} {\bibinfo  {journal} {Journal of Energy Chemistry}\ }\textbf
		{\bibinfo {volume} {94}},\ \bibinfo {pages} {577} (\bibinfo {year}
		{2024})}\BibitemShut {NoStop}%
\end{thebibliography}
%apsrev4-2.bst 2019-01-14 (MD) hand-edited version of apsrev4-1.bst
%Control: key (0)
%Control: author (8) initials jnrlst
%Control: editor formatted (1) identically to author
%Control: production of article title (0) allowed
%Control: page (0) single
%Control: year (1) truncated
%Control: production of eprint (0) enabled
%

\end{document}